\documentclass[lettersize,journal]{IEEEtran}
\usepackage{amsmath,amssymb,amsfonts}
\usepackage{algorithmic}
\usepackage{listings}
\usepackage{graphicx}
\usepackage{tabularx}
\usepackage{color, colortbl}
\usepackage{xurl}
\usepackage{multirow}
\usepackage{url}
\usepackage{hyperref}
\usepackage{textcomp}
\usepackage{booktabs}
\usepackage{enumitem}
\usepackage{ulem}
\usepackage{adjustbox}
\usepackage{pifont} % For symbols like checkmarks
\usepackage{subcaption} % 解决子图编号问题，避免使用 subfigure（已过时）

\usepackage[table,xcdraw]{xcolor} % 先加载 xcolor 包，并指定所有需要的选项
\usepackage{geometry}
\usepackage[most]{tcolorbox}

\usepackage{threeparttable}
\usepackage{tikz}

\definecolor{codegreen}{rgb}{0,0.6,0}
\definecolor{codegray}{rgb}{0.5,0.5,0.5}
\definecolor{codepurple}{rgb}{0.58,0,0.82}
\definecolor{backcolour}{rgb}{0.95,0.95,0.92}

\definecolor{main}{HTML}{5989cf}    % setting main color to be used
\definecolor{sub}{HTML}{cde4ff}     % setting sub color to be used

\newcommand{\circledwhite}[1]{%
  \tikz[baseline=(char.base)]{
    \node[circle,fill=black,inner sep=1pt] (char) {\textcolor{white}{\sffamily\bfseries #1}};
  }%
}

\tcbset{
    sharp corners,
    colback = white,
}                           % setting global options for tcolorbox

\newtcolorbox{boxA}{
    boxrule = 1.5 pt,
    colframe = black % frame color
}

\lstdefinestyle{mystyle}{
    backgroundcolor=\color{white},   
    commentstyle=\color{blue},
    keywordstyle=\color{red},
    numberstyle=\tiny\color{codegray},
    stringstyle=\color{codepurple},
    basicstyle=\ttfamily\footnotesize,
    breakatwhitespace=false,         
    breaklines=true,                 
    captionpos=b,                    
    keepspaces=true,                 
    numbers=left,                    
    numbersep=0pt,    
    showspaces=false,                
    showstringspaces=false,
    showtabs=false,  
    tabsize=1
}

\def\BibTeX{{\rm B\kern-.05em{\sc i\kern-.025em b}\kern-.08em
    T\kern-.1667em\lower.7ex\hbox{E}\kern-.125emX}}

\usepackage{listings}
\usepackage{xcolor}

\definecolor{codeblue}{rgb}{0.0, 0.1, 0.6}   % Keywords: 深蓝
\definecolor{codered}{rgb}{0.6, 0.0, 0.0}    % Strings: 深红
\definecolor{codegreen}{rgb}{0.0, 0.45, 0.0} % Comments: 墨绿
\definecolor{codegray}{rgb}{0.2, 0.2, 0.2}   % Default text

\begin{document}
%
% paper title
% Titles are generally capitalized except for words such as a, an, and, as,
% at, but, by, for, in, nor, of, on, or, the, to and up, which are usually
% not capitalized unless they are the first or last word of the title.
% Linebreaks \\ can be used within to get better formatting as desired.
% Do not put math or special symbols in the title.
%\title{\textsc{DefendCLI}: Command-Line Driven Real-Time Attack Provenance Construction}
\title{\huge \textsc{DefendCLI}: \{Command-Line\} Driven Attack Provenance Examination}

\author{Peilun Wu\IEEEauthorrefmark{1}\IEEEauthorrefmark{2}, Nan Sun\IEEEauthorrefmark{1}, Nour Moustafa\IEEEauthorrefmark{1}, Youyang Qu\IEEEauthorrefmark{2}, Ming Ding\IEEEauthorrefmark{2}\\
\IEEEauthorrefmark{1}UNSW at the Australian Defence Force Academy (UNSW Canberra), Australia\\ 
\IEEEauthorrefmark{2}Commonwealth Scientific and Industrial Research Organisation (CSIRO Data61), Australia\\
\thanks{Manuscript received [Month] [Day], [Year]; revised [Month] [Day], [Year]. (Optional: This work was supported by the CSIRO Data61 Scholarship / Australian Research Council under Grant XXXXX.) \textit{(Corresponding author: Peilun Wu.)}}
\thanks{Peilun Wu is with the School of Engineering and Technology, UNSW Canberra, ACT 2600, Australia, and also with CSIRO Data61, Australia (e-mail: peilun.wu@student.unsw.edu.au).}
\thanks{Nan Sun and Nour Moustafa are with the School of Engineering and Technology, UNSW Canberra, ACT 2600, Australia (e-mail: nan.sun@unsw.edu.au; nour.moustafa@unsw.edu.au).}
\thanks{Youyang Qu and Ming Ding are with CSIRO Data61, Australia (e-mail: youyang.qu@data61.csiro.au; ming.ding@data61.csiro.au).}}

\markboth{Journal of \LaTeX\ Class Files,~Vol.~18, No.~9, September~2020}%
{How to Use the IEEEtran \LaTeX \ Templates}

% make the title area
\maketitle

% As a general rule, do not put math, special symbols or citations
% in the abstract
\begin{abstract}
%-------------------------------------------------------------------------------
%In the paper, we introduce \textsc{DefendCLI}, a predominantly command-line-driven endpoint security solution, which is intended to improve the detection and response capabilities of Endpoint Detection and Response (EDR) systems.
Endpoint Detection and Response (EDR) solutions embrace the method of attack provenance graph to discover unknown threats through system event correlation.
However, this method still faces some unsolved problems in the fields of interoperability, reliability, flexibility, and practicability to deliver actionable results.
Our research highlights the limitations of current solutions in detecting obfuscation, correlating attacks, identifying low-frequency events, and ensuring robust context awareness in relation to command-line activities.

To address these challenges, we introduce \textsc{DefendCLI},  an innovative system leveraging provenance graphs that, for the first time, delves into command-line-level detection.
By offering finer detection granularity, it addresses a gap in modern EDR systems that has been overlooked in previous research. 
Our solution improves the precision of the information representation by evaluating differentiation across three levels: unusual system process calls, suspicious command-line executions, and infrequent external network connections. 
This multi-level approach enables EDR systems to be more reliable in complex and dynamic environments. 

Our evaluation demonstrates that \textsc{DefendCLI} improves precision by approximately $1.6 \times$ compared to the state-of-the-art methods on the DARPA Engagement Series attack datasets. 
Extensive real-time industrial testing across various attack scenarios further validates its practical effectiveness. 
The results indicate that \textsc{DefendCLI} not only detects previously unknown attack instances, which are missed by other modern commercial solutions, but also achieves a $2.3 \times$ improvement in precision over the state-of-the-art research work.

%Traditional methods construct attack provenance graphs focused on the flow of information using system events for anomaly detection, but they often include excessive and irrelevant information, making security analysis more complex.
%\textsc{DEFENDCLI} refines attack provenance graphs through partial command-line differentiation analysis from interactive shell executions and application operations, simplifying graph structure and significantly improving the precision of information representation.
%Our evaluation shows that \textsc{defendcli} increases the precision by 60\% compared to current solutions, successfully detecting 14 malicious activities in the E3 benchmark datasets. These results demonstrate \textsc{defendcli}'s superior capability in detecting cyber threats in real-world scenarios.
%Furthermore, \textsc{DefendCLI} undergoes an extensive real-time evaluation in various attack scenarios to verify its practicability and effectiveness against commercial EDR systems. The findings indicate that \textsc{DefendCLI} outperforms current alternatives, identifying 65\% more attacks that other existing solutions fail to detect.
\end{abstract}


\section{Introduction}
\label{sec:introduction}
%The constantly changing threat landscape demands advanced solutions capable of efficiently identifying and counteracting complex attacks.
Traditional security measures, like anti-virus software and firewalls, frequently fall short in detecting contemporary threats that employ advanced evasion strategies, such as memory injection and fileless attacks.
These limitations highlight the need for advanced Endpoint Detection and Response (EDR) solutions to detect complex and stealthy attacks, providing better insights into behavior and operations to improve security incident detection.
An essential improvement of modern EDR systems is the attack provenance graph, which visually represents the relationships between different system events to discover potential advanced threats.

While there is growing interest in provenance-based EDR systems, their adoption in real-world scenarios remains limited, and comprehensive evaluation is still lacking.
To address this gap, we deployed a provenance-based EDR \cite{wang2020you} to automatically analyze system audit logs from the Microsoft Cyber Defense Operations Center (CDOC) and passed the results to security analysts for verification.
During this process, analysts encountered challenges in efficiently utilizing the provenance-based results and we summarized four key unresolved research questions (RQs) that require further investigation.

\begin{figure}[t]
    \centering
    \includegraphics[width=.9\linewidth]{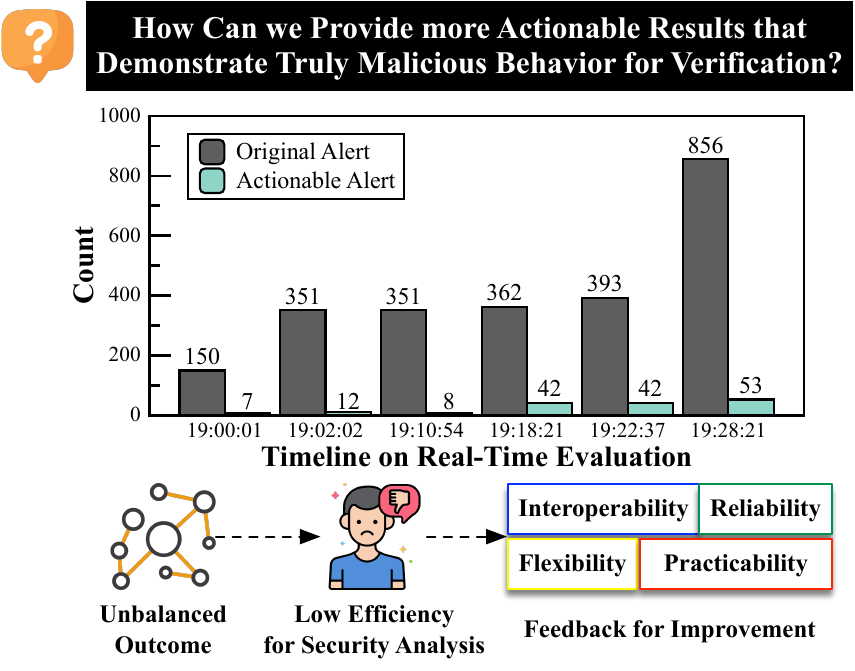}
    \caption{\textbf{Motivation Feedback:}  A provenance-based detector \cite{wang2020you} deployed in an enterprise environment. Only 0.07\% of the original alerts were truly actionable, providing accurate attack-related malicious activities, such as command-line executions and network connections, for a valid verification.}
     %Security analysts recommend improvements in four key areas.}
    \label{fig:action}
    %\vspace{-15pt} % 调整图像与下文之间的距离
\end{figure}

\circledwhite{1} \textbf{\uline{RQ1-Interoperability}: How can we reduce the complexity of attack provenance graphs to provide more actionable results?} 
Provenance graphs are often overly complex and cluttered with invalid information, making interpretation difficult and misaligned with analysts' priorities. 

\begin{figure}[t]
    \centering
    \includegraphics[width=.9\linewidth]{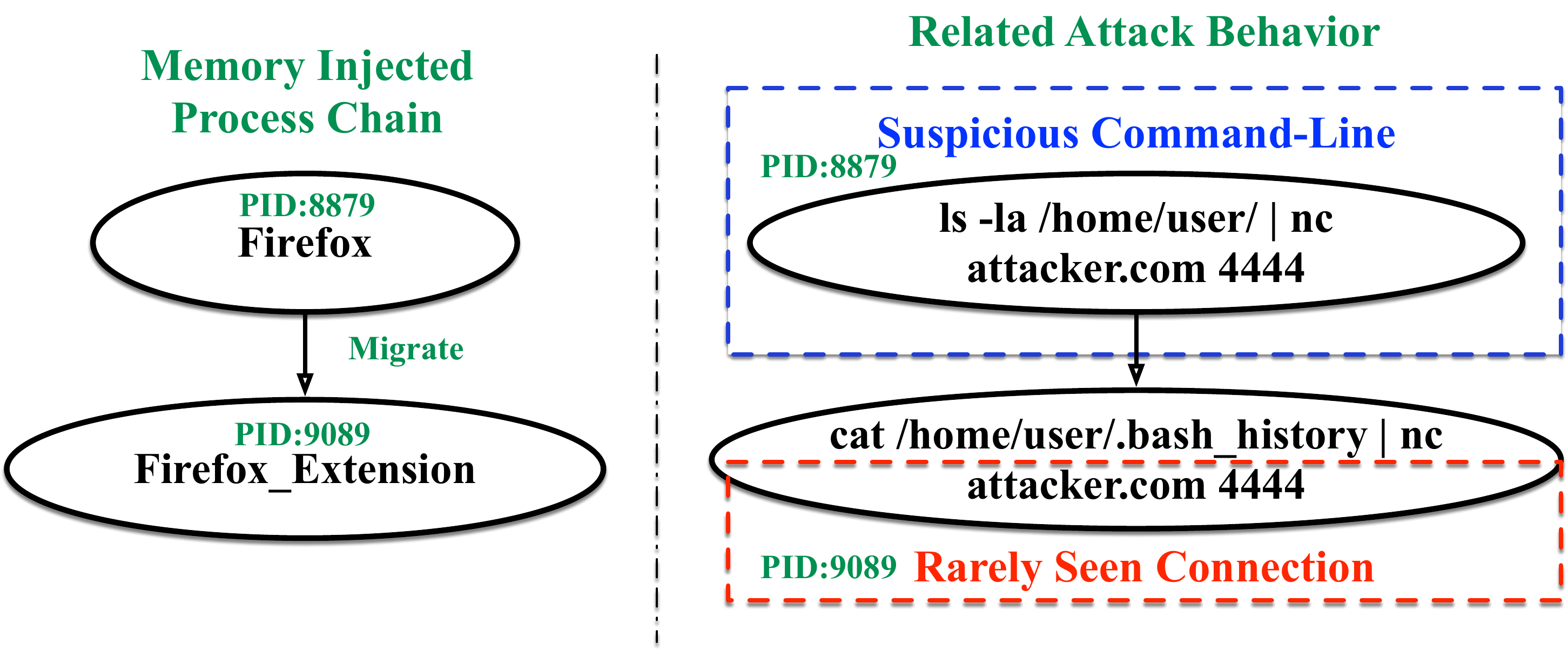}
    \caption{\textbf{Omission:} Even if a process chain is malicious, but it cannot be analyzed if we do not known what command-line are executed, of which effective threat verification is affected. }
    \label{fig:p1}
    %\vspace{-15pt} % 调整图像与下文之间的距离
\end{figure}

Malicious activities often blend seamlessly into legitimate system operations, causing provenance-based EDRs to generate an overwhelming number of anomalies that lack actionable insights, as illustrated in Fig \ref{fig:action} and Fig \ref{fig:p1}. 
While these results may not necessarily be false alarms, they fail to provide definitive evidence of an intrusion, making manual investigation even more challenging and hindering efforts to trace the root cause and focus on suspicious activities.

These challenges highlight the need for a more refined approach to provenance graph construction, which can minimize complexity, enhance detection accuracy, and deliver more relevant, actionable insights for security analysts.

 \circledwhite{2} \textbf{\uline{RQ2-Reliability}: How can we provide a more effective approach for anomaly detection on different operating system platforms?}
Existing anomaly detection methods \cite{hassan2020tactical, wang2022threatrace, hassan2018towards} often adopt a \textit{one-size-fits-all} approach, failing to account for OS-specific characteristics. 
While this generalization increases versatility, it overlooks critical real-world complexities.

Enterprise security policies often segment networks into distinct zones, such as the demilitarized zone (DMZ) and internal office networks. 
The DMZ primarily hosts externally facing services, like DNS and web servers that typically run Linux, whereas internal office networks support end-user devices, predominantly Windows-based.
Each segment presents unique behavioral patterns, attack surfaces, and vulnerabilities, making OS-specific anomaly detection essential for effective threat identification.

A more tailored approach, one that considers the distinct operational norms and security risks of each OS, can significantly improve the accuracy and efficiency of EDR solutions.
By refining detection mechanisms to align with platform-specific behaviors, EDR systems can reduce false positives, enhance threat visibility, and provide more actionable results.

%\begin{figure*}
%    \centering
 %   \includegraphics[width=\linewidth]{motivation_hhh.pdf}
%    \caption{\uline{Comparison Between Traditional Method and \textsc{DefendCLI} Focused Structure}: 
%    The box of \textcolor{blue}{$\blacksquare$} indicates InfoPath, where \#1 is malicious and \#2 is normal. 
%    The box of \textcolor{red}{$\blacksquare$} indicates \textsc{DefendCLI} focused structure, which is the truly malicious part within the entire box of \textcolor{black}{$\blacksquare$} provenance graph.}
%    \label{fig:overview}
    %\vspace{-15pt} % 调整图像与下文之间的距离
%\end{figure*}

%\subsection{RQ3: Tailor to Conditions}
 \circledwhite{3} \textbf{\uline{RQ3-Flexibility}: How can we tailor AI-based detection mechanisms to minimize false positives and improve adaptability in dynamic environments?}   
The adoption of AI in security detection and response is accelerating due to its superior capability in identifying zero-day threats compared to traditional signature-based systems.

However, AI-driven solutions introduce significant challenges, particularly an elevated false positive rate in dynamic environments where factors such as new device deployments, software updates, and network reconfigurations frequently trigger erroneous alerts.  

Moreover, the inherent ``black-box" nature of AI complicates security investigations, as opaque decision-making processes undermine interpretability, trust, and operational transparency.

To address these limitations, AI models should incorporate adaptive learning mechanisms that accommodate environmental variability while maintaining a stringent false positive control. Enhancing explainability through interpretable AI techniques and leveraging contextual awareness in anomaly detection can further improve precision, thereby fostering greater trust and operational efficacy in security operations.

%\subsection{RQ5: Moving Beyond Binary Decisions}
\circledwhite{4} \textbf{\uline{RQ4-Practicability}: How can we evolve beyond binary decision-making frameworks in EDR systems to enable more granular and nuanced threat analysis?} 
Traditional security frameworks that classify alerts as strictly malicious or benign are increasingly inadequate, as modern cyber threats often blur the lines between legitimate and malicious activity. Advanced attack techniques exploit these gray areas, rendering binary classifications ineffective in identifying sophisticated threats.  

To address this limitation, AI-driven security solutions are shifting toward threat-scoring models that assess incidents based on severity, providing a more nuanced analysis. However, many of these systems remain rule-based, necessitating frequent updates to adapt to evolving attack patterns. This continuous maintenance imposes a significant operational burden on security vendors, limiting scalability and responsiveness to emerging threats.  

To enhance efficiency, a more automated, adaptive and actionable security orchestration framework is required - one that minimizes manual intervention while dynamically refining threat assessments. By integrating machine learning-driven anomaly detection, contextual analysis, and automated incident response, EDR systems can improve detection precision, reduce operational overhead, and offer a more intelligent, scalable approach to threat management.

%\subsection{Our Contributions}
\textbf{\uline{Our Contributions.}} 
To address the aforementioned RQs, \textsc{DefendCLI} presents a practical solution that, for the first time, delves into the command-line-level detection within the attack provenance graph, improving accuracy while avoiding unrelated information.

Command-line interactions encompass both \textbf{interactive shell operations and OS API executions}, providing direct visibility into system-level behaviors.
%Although attackers may attempt partial command-line obfuscation or escapes, as the complexity and number of attack steps increase, many system functions and applications require further executions, which inevitably generate additional command-line activities that cannot be entirely concealed.
Despite attackers' attempts at command-line obfuscation, sophisticated multi-stage attacks require numerous system function executions, inevitably producing traceable command-line artifacts that cannot be fully concealed from detection systems.
Despite the widespread adoption of command-line monitoring in rule-based EDR systems, its formal representation in attack provenance graphs and systematic integration with AI-driven anomaly detection frameworks remains underdeveloped.

Our solution employs an explainable AI system to match the known malicious command-line with well-defined rule sets and detect unknown command-line activities with the SimHash-Based \cite{sadowski2007simhash} anomaly detection mechanism.
Furthermore, the overall solution focuses on multi-level attack provenance examination, including unusual system process calls, suspicious command-line executions, and infrequent external network connections, which can resist defense evasion to some extent.

To provide a more automated security orchestration scheme, we integrate Large Language Models (LLMs) into our threat recommendation system to generate rapid preliminary assessments, enabling efficient triage before comprehensive threat analysis.
Extensive experiments and evaluations show that \textsc{DEFENDCLI} effectively manages real-world cyber threats by identifying previously undetected attack instances.

Our contributions are summarized as follows:

\begin{itemize}[leftmargin=*]
    \item \textbf{\uline{Interoperability:}} 
    We introduce a novel attack provenance graph structure specifically tailored for command-line-driven, multi-level differentiation detection. This refined structure reduces the complexity associated with traditional provenance graphs, resulting in more precise information representation and improved actionable insights for security analysts. 
    \item \textbf{\uline{Reliability:}}
    Our approach incorporates an explainable anomaly detection framework, integrating defense evasion signatures and known attack patterns from both Linux and Windows platforms into an AI-driven architecture. The system also employs graph algorithms for multi-level provenance analysis alongside SimHash-based differentiation of unknown command-line executions, significantly reducing false positives by moving away from pre-training unreliable normal behavior profiles.
    \item \textbf{\uline{Flexibility:}} 
    The framework supports dynamic adaptation through configurable hyperparameters, optimizing performance for various operational environments. By leveraging a retrieval-augmented generation (RAG) approach with LLMs, we generate prioritized incident assessments with detailed explanations, named snapshot alarms, thus streamlining alert triage and mitigating analyst alert fatigue.
    \item \textbf{\uline{Practicability:}} 
     We evaluate the performance of our system against state-of-the-art research work and modern commercial EDR tools on academic benchmark datasets and industrial real-time detection.
     The result shows the practical applicability and superior effectiveness of our solution for modern offensive and defensive confrontation.
\end{itemize}

%\begin{figure}
 %   \centering
  %  \includegraphics[width=.9\linewidth]{comparison_new.pdf}
    %\caption{In the analysis of the $F_1$ scores, detection systems were analyzed using the DARPA Transparent Computing Engagements 3-5 datasets\cite{TCP}. Given the absence of a broad comparative study, our objective is to succinctly present the findings, highlighting the minimum scores reported in the original studies, to determine the overall direction of the scholarly research.}
   % \caption{$F_1$ score compared on the E1 - E5 datasets\cite{TCP}.}
%\label{fig:comparison}
%\end{figure}

\section{Related Work}
\label{sec:related-work}

This section provides a comparative analysis of related work across four key dimensions critical for command-line activity detection: \textbf{obfuscation detection (OD), attack correlation (AC), low-frequency detection (LFD), and context awareness (CA)} \cite{csi}.
Table~\ref{tab:detection_level} presents a feature analysis of \textsc{DefendCLI} relative to prior methods based on their algorithmic designs (AD).

\circledwhite{1} \textbf{\uline{OD: Challenges in Obfuscation Detection}} \cite{goyal2023sometimes}.
Most existing approaches primarily leverage structural anomalies, a strategy that may not fully capture the semantics of obfuscated commands (e.g., polymorphic commands or encoded payloads). \textsc{Holmes} \cite{milajerdi2019holmes} provides effective detection via TTP-based rules. Methods like \textsc{Unicorn} \cite{han2020unicorn}, \textsc{DepComm} \cite{xu2022depcomm}, \textsc{ProGrapher} \cite{yang2023prographer}, \textsc{Nodlink} \cite{li2023nodlink}, and \textsc{DISTDET} \cite{dong2023distdet} focus largely on structural changes or statistical deviations. While effective for general anomalies, these systems face challenges in parsing sophisticated obfuscation techniques without semantic analysis.

\circledwhite{2} \textbf{\uline{AC: Granularity in Attack Correlation}} \cite{zipperle2022provenance}.
Effective correlation is essential for reconstructing attack chains. \textsc{Nodlink} \cite{li2023nodlink} demonstrates strong performance by employing Steiner trees for node-level correlation, while \textsc{Holmes} \cite{milajerdi2019holmes} utilizes TTP rules to provide a cohesive view of adversarial behavior. \textsc{Unicorn}, \textsc{DepComm}, and \textsc{ProGrapher} rely on dependency tracking, which captures structural relationships. \textsc{DISTDET} \cite{dong2023distdet} emphasizes efficient log reduction and deduplication, which prioritizes scalability but may limit the resolution of complex relationships in multi-stage APTs.

\circledwhite{3} \textbf{\uline{LFD: Sensitivity to Low-Frequency Events}} \cite{mukherjee2023evading}.
Detecting rare, non-repetitive patterns is vital for identifying novel or targeted attacks. Many current methods (\textsc{Unicorn}, \textsc{DepComm}, \textsc{ProGrapher}, \textsc{Nodlink}, \textsc{DISTDET}) utilize historical frequency profiles, which are effective for repetitive behaviors but may be less sensitive to rare events lacking precedence. \textsc{Holmes} \cite{milajerdi2019holmes} offers detection through predefined rules, whereas anomaly-based baselines in other systems must balance sensitivity with false-positive rates when encountering stealthy, low-frequency threats.

\circledwhite{4} \textbf{\uline{CA: Context Awareness Integration}} \cite{li2021threat}.
Distinguishing benign from malicious commands often depends on execution context (e.g., user privileges, parent processes). \textsc{DepComm} \cite{xu2022depcomm} incorporates system attributes as dependencies to enhance context. \textsc{Holmes} \cite{milajerdi2019holmes} uses rule-based context for known attacks. Approaches such as \textsc{Unicorn}, \textsc{ProGrapher}, \textsc{Nodlink}, and \textsc{DISTDET} primarily focus on graph structure. Integrating robust environmental variables remains a key area for development to accurately differentiate between legitimate operations and malicious mimicry.

\begin{figure*}
    \centering
    \includegraphics[width=\linewidth]{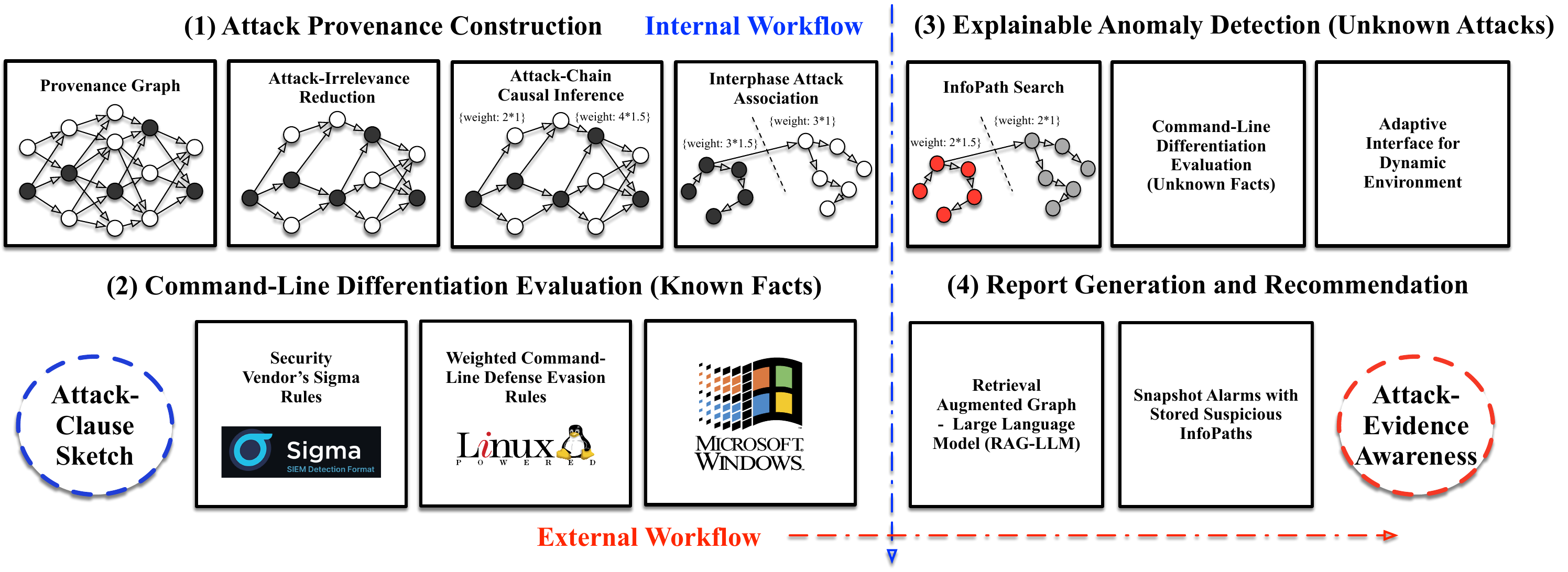}
    \caption{\textbf{Attack-Clause Sketch} and \textbf{Attack-Evidence Awareness}:  
The \textbf{attack-clause sketch} module includes (1) \textbf{provenance construction} for building behavior graphs and (2) \textbf{command-line evaluation} to refine weights using known attacks. The \textbf{attack-evidence awareness} module handles (3) \textbf{anomaly detection} for unknown threats and (4) \textbf{reporting} via RAG-LLM.
}
    \label{fig:DefendCLI}
    %\vspace{-15pt} 
\end{figure*}

\begin{table}[t]
\caption{Feature Analysis of \textsc{DEFENDCLI} Relative to Prior Work. 
\textcolor{black}{\ding{108}}: Comprehensive, \textcolor[HTML]{7F7F7F}{\ding{108}}: Substantial, \textcolor[HTML]{B3B3B3}{\ding{108}}: Partial, \textcolor[HTML]{000000}{\ding{109}}: Limited.}
\centering
\begin{tabular}{@{}l|ccccc@{}}
\toprule
\textbf{Method}     & \textbf{OD} & \textbf{AC} & \textbf{LFD} & \textbf{CA} &\textbf{AD}\\ \midrule
\textsc{Holmes} \cite{milajerdi2019holmes}     
& \textcolor[HTML]{7F7F7F}{\ding{108}} % Substantial
& \textcolor[HTML]{7F7F7F}{\ding{108}} % Substantial
& \textcolor[HTML]{B3B3B3}{\ding{108}} % Partial
& \textcolor[HTML]{B3B3B3}{\ding{108}} % Partial
& ATT\&CK Framework
\\
\midrule
\textsc{Unicorn} \cite{han2020unicorn}     
& \textcolor[HTML]{000000}{\ding{109}} % Limited
& \textcolor[HTML]{7F7F7F}{\ding{108}} % Substantial
& \textcolor[HTML]{000000}{\ding{109}} % Limited
& \textcolor[HTML]{000000}{\ding{109}} % Limited
& Path Traversal
\\
\midrule
\textsc{DepComm} \cite{xu2022depcomm}   
& \textcolor[HTML]{000000}{\ding{109}} % Limited
& \textcolor[HTML]{7F7F7F}{\ding{108}} % Substantial
& \textcolor[HTML]{000000}{\ding{109}} % Limited
& \textcolor[HTML]{7F7F7F}{\ding{108}} % Substantial
&  Graph Partition
\\
\midrule
\textsc{Prographer} \cite{yang2023prographer} 
& \textcolor[HTML]{000000}{\ding{109}} % Limited
& \textcolor[HTML]{7F7F7F}{\ding{108}} % Substantial
& \textcolor[HTML]{000000}{\ding{109}} % Limited
& \textcolor[HTML]{000000}{\ding{109}} % Limited
&  Graph Partition
\\
\midrule
\textsc{Nodlink} \cite{li2023nodlink}   
& \textcolor[HTML]{000000}{\ding{109}} % Limited
& \textcolor{black}{\ding{108}}  % Comprehensive
& \textcolor[HTML]{000000}{\ding{109}} % Limited
& \textcolor[HTML]{000000}{\ding{109}} % Limited
& Steiner Tree Partition
\\
\midrule
\textsc{DISTDET} \cite{dong2023distdet}      
& \textcolor[HTML]{000000}{\ding{109}} % Limited
& \textcolor[HTML]{000000}{\ding{109}} % Limited
& \textcolor[HTML]{000000}{\ding{109}} % Limited
& \textcolor[HTML]{000000}{\ding{109}} % Limited
& Graph Partition
\\
\midrule
\textsc{DefendCLI}   
& \textcolor{black}{\ding{108}} % Comprehensive
& \textcolor{black}{\ding{108}} % Comprehensive
& \textcolor{black}{\ding{108}} % Comprehensive
& \textcolor{black}{\ding{108}} % Comprehensive
& Multi-Level Deviation
\\
\bottomrule
\end{tabular}
\label{tab:detection_level}
\end{table}

\section{\textsc{DefendCLI}}
\label{sec:system_design}

The architecture of \textsc{DefendCLI}, illustrated in Figure \ref{fig:DefendCLI}, adopts a highly modular design to ensure extensibility and clarity in addressing the challenges of command-line attack detection. The system is fundamentally composed of two sequential computational phases:
\textbf{Attack-Clause Sketch}, dedicated to the construction, refinement, and structural quantification of the provenance graph; and \textbf{Attack-Evidence Awareness}, responsible for semantic analysis, anomaly detection, and explainable threat reporting.

\subsection{Attack-Clause Sketch: Graph Construction and Quantification}
This module focuses on transforming raw system telemetry into a structured, quantifiable representation suitable for causal analysis. It involves graph refinement, probabilistic node scoring, and interphase community association.

\begin{figure*}[t]
    \centering
    \includegraphics[width=\linewidth]{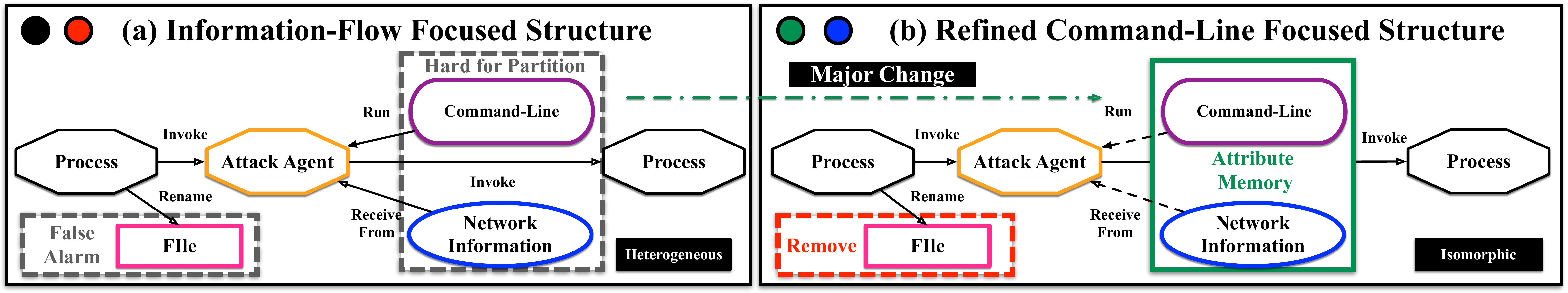}
    \caption{Comparison of (a) \textbf{Information-Flow Focused Structure} (heterogeneous nodes) and (b) \textbf{Refined Command-Line Focused Structure} (isomorphic process nodes with attributes). The refined structure reduces graph noise by encapsulating command and network data within process nodes.}
    \label{fig:graph_construction}
\end{figure*}

\begin{figure*}[t]
    \centering
    \includegraphics[width=\linewidth]{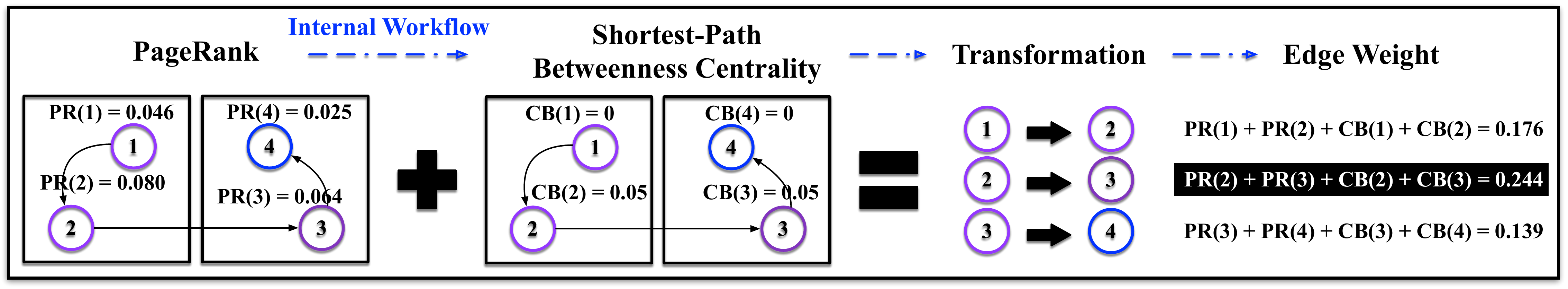}
    \caption{\textbf{Attack-chain causal inference}: Node importance is calculated using PageRank (for rarity) and Shortest-Path Betweenness Centrality (for structural bridges). These scores are aggregated to determine initial edge weights.}
    \label{fig:weight_assignment}
\end{figure*}

\begin{figure}[t]
    \centering
    \includegraphics[width=\linewidth]{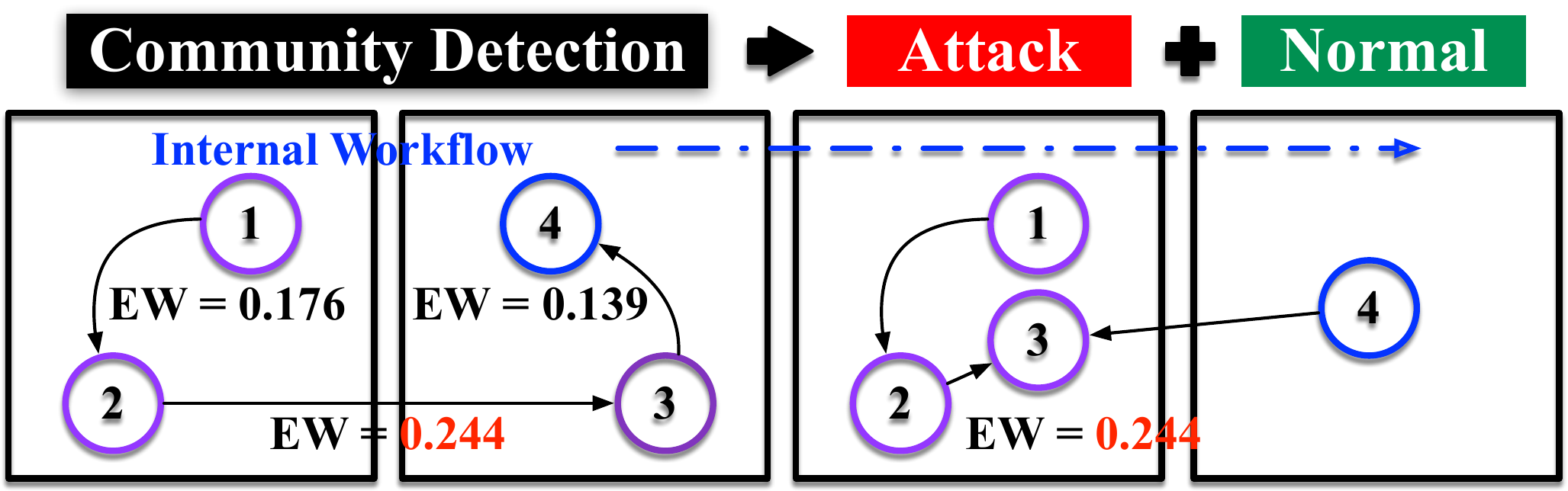}
    \caption{\textbf{Interphase attack association}: The Leiden community detection algorithm groups related process nodes, overcoming fragmentation issues to correlate multi-stage attacks.}
    \label{fig:interphase}
\end{figure}

\subsubsection{Refined Provenance Structure and Reduction}
Traditional approaches often construct heterogeneous graphs where processes, files, sockets, and command strings are distinct nodes (Figure \ref{fig:graph_construction}(a)). This results in high-dimensional, noisy graphs where semantic context is diluted.

To address this, we define a \textit{Refined Command-Line Focused Structure} (Figure \ref{fig:graph_construction}(b)). Let the system execution be represented as a directed graph $G = (V, E, \mathcal{A})$. Here, $V$ represents the set of uniform system process nodes. $E \subseteq V \times V$ denotes direct causal dependencies (e.g., process forking). Crucially, command-line arguments and network contexts are not separate nodes but are encapsulated as an attribute set $\mathcal{A}_v = \{\text{cmd}_v, \text{net}_v, \dots\}$ resident within each process node $v \in V$. This isomorphic structure preserves semantic richness while significantly reducing graph complexity for downstream embedding and traversal algorithms.

To mitigate the dependency explosion problem \cite{hossain2020combating}, we apply a two-stage graph reduction function $\Phi: G \to G'$.

First, \textbf{Semantic Pruning} removes nodes lacking command-line relevance. We define a subgraph $G_{sub} \subset G$ as irrelevant if for all nodes $v \in G_{sub}$, the command attribute $\text{cmd}_v = \emptyset$. These subgraphs are pruned as they typically represent benign background noise.

Second, \textbf{Acyclic Transformation} ensures valid causal tracing by eliminating cyclic dependencies. $G'$ is transformed into a Directed Acyclic Graph (DAG) via depth-first search based cycle removal, ensuring that any traversal represents a valid temporal flow of events.

\subsubsection{Causal Inference via Hybrid Node Scoring}
Identifying critical nodes within massive provenance graphs requires quantifying their structural importance. We employ a hybrid scoring mechanism combining probabilistic ranking and centrality measures to determine initial node weights, as depicted in Figure \ref{fig:weight_assignment}.

\textbf{Quantifying Rareness (Inverted PageRank).} We adapt the PageRank algorithm \cite{bianchini2005inside} to identify anomalous nodes based on topological scarcity. Let $PR(v)$ denote the PageRank score of node $v$. The iterative computation is defined as:
\begin{equation}
PR(v)_{t+1} = (1 - d) + d \sum_{u \in \text{In}(v)} \frac{PR(u)_t}{L(u)}
\end{equation}
where $d \approx 0.85$ is the damping factor, $\text{In}(v)$ are incoming predecessors, and $L(u)$ is the out-degree of $u$. Unlike standard applications where high PR indicates prominence, in attack provenance, nodes with low in-degree (lower PR) often signify rare, potentially malicious execution paths. We normalize these scores to emphasize rarity.

\textbf{Identifying Root Causes (Betweenness Centrality).} To find pivotal "bridge" nodes that connect disparate attack phases (e.g., initial access leading to privilege escalation), we utilize Shortest-Path Betweenness Centrality ($CB$) \cite{brandes2001faster}. For a node $v$, $CB(v)$ quantifies the fraction of shortest paths between all node pairs $(s,t)$ that pass through $v$:
\begin{equation}
CB(v) = \sum_{s \neq v \neq t} \frac{\sigma_{st}(v)}{\sigma_{st}}
\end{equation}
where $\sigma_{st}$ is the total number of shortest paths from $s$ to $t$, and $\sigma_{st}(v)$ is the count of those paths passing through $v$. High $CB$ scores highlight structurally critical bottlenecks.

\textbf{Edge Weight Initialization.} To facilitate path-based analysis, these node-level metrics must be projected onto the edges. For any directed edge $e_{mn} \in E'$ connecting parent $m$ to child $n$, we define the initial Base Edge Weight ($W_{mn}^{base}$) as the normalized aggregate of the structural significance of both connected nodes:
\begin{equation}
W_{mn}^{base} = \text{Norm}\left( PR(m) + PR(n) + CB(m) + CB(n) \right)
\end{equation}
where $\text{Norm}(\cdot)$ scales the resulting magnitude to the range $[0,1]$. This weight represents the inherent structural interest of the transition from $m$ to $n$.

\subsubsection{Interphase Attack Association via Community Detection}
Attacks often manifest as distinct phases separated by time, making causal linkage difficult. To correlate these multi-stage behaviors, we partition the reduced graph $G'$ into semantically coherent clusters.

We utilize the Leiden algorithm \cite{leidenalg}, an improvement over Louvain that guarantees connected communities and optimizes modularity. As illustrated in Figure \ref{fig:interphase}, Leiden partitions the graph into a set of disjoint communities $\mathcal{C} = \{C_1, C_2, \dots, C_k\}$.
Each detected community $C_k$ is assigned a \textit{Community Score} ($S_{C_k}$) based on the aggregate risk density within that cluster. This score serves as a coherence multiplier in subsequent analysis, reinforcing links between activities occurring within the same attack phase context.

\subsection{Attack-Evidence Awareness: Detection and Reporting}
\begin{figure*}
    \centering
    \includegraphics[width=\linewidth]{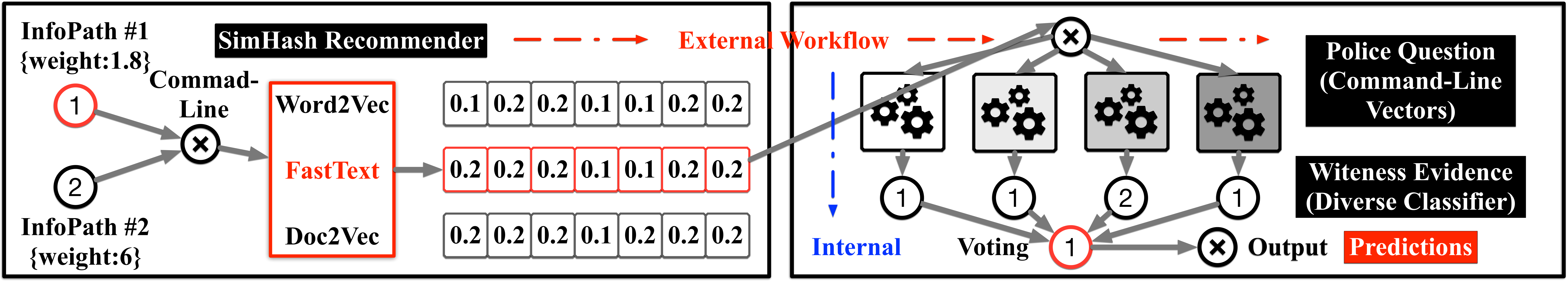}
    \caption{\textbf{Explainable anomaly detection mechanism}: The process involves retrieving ranked InfoPaths, analyzing command-line differences via SimHash, and using a multi-classifier "Police-Witness" voting policy for final verdict.}
    \label{fig:detection}
\end{figure*}

The second phase leverages the weighted provenance graph to detect threats. This involves differentiating command-line semantics against known signatures and employing ensemble learning for unknown anomaly detection, as outlined in Figure \ref{fig:detection}.

\subsubsection{Differentiation of Known Threats (Rule-Based Boosting)}
To integrate expert knowledge, we evaluate process command-lines against curated threat intelligence, specifically Sigma rules \cite{sigma} and defense evasion patterns \cite{redcanary}.
Let $R(e_{mn})$ be a boolean indicator function that returns 1 if the command-line context $\text{cmd}_n$ associated with edge $e_{mn}$ matches a known threat signature, and 0 otherwise. We introduce a Risk Coefficient $\alpha > 1$ corresponding to the severity of the matched rule.
The base edge weight is updated to a \textit{Refined Edge Weight} ($W_{mn}^{refined}$) to reflect this deterministic knowledge:
\begin{equation}
W_{mn}^{refined} = \begin{cases} 
W_{mn}^{base} \times \alpha & \text{if } R(e_{mn}) = 1 \\
W_{mn}^{base} & \text{otherwise}
\end{cases}
\end{equation}
This operation significantly amplifies the weights of paths containing known malicious indicators, ensuring their prioritization during retrieval.

\subsubsection{Differentiation of Unknown Anomalies (SimHash Ensemble)}
Detecting novel attacks requires identifying subtle deviations in command-line semantics, often disguised through obfuscation. We propose an approach combining semantic embedding with locality-sensitive hashing.

\begin{figure}[t]
    \centering
    \includegraphics[width=\linewidth]{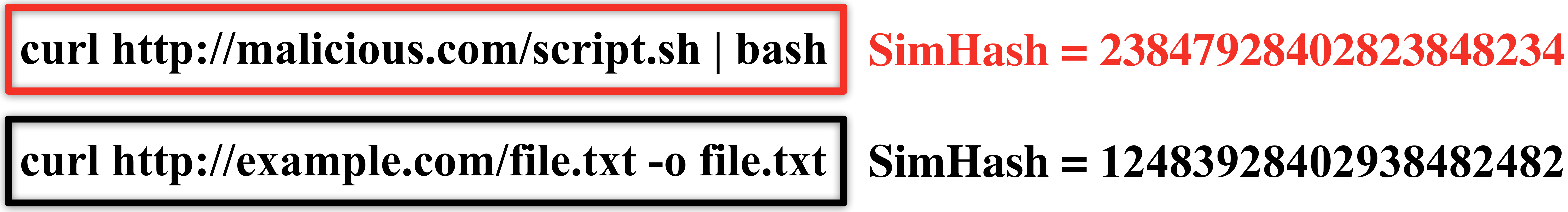}
    \caption{Example of SimHashed command-line executions. SimHash maps semantically similar commands to nearby binary signatures, allowing detection of obfuscated variations (red box) via Hamming distance.}
    \label{fig:simhash}
\end{figure}

\textbf{Semantic Hashing via SimHash.}
Standard embeddings (e.g., Word2Vec) result in high-dimensional vectors where similarity search is computationally expensive. To efficiently detect obfuscated variations, we apply SimHash \cite{sadowski2007simhash}.
Let $\mathbf{v}_c \in \mathbb{R}^d$ be the dense vector representation of a command string $c$. The SimHash function $H: \mathbb{R}^d \to \{0,1\}^f$ maps this vector to an $f$-bit binary signature $h_c$. A crucial property of SimHash is that the Hamming distance between two signatures approximates the cosine similarity of their original vectors:
\begin{equation}
\text{Hamming}(h_{c_i}, h_{c_j}) \approx \theta(\mathbf{v}_{c_i}, \mathbf{v}_{c_j})
\end{equation}
This allows us to efficiently quantify semantic deviation, as shown in Figure \ref{fig:simhash}, where obfuscated commands yield small but detectable Hamming distances from benign counterparts.

\textbf{Ensemble "Police-Witness" Voting.}
To avoid algorithmic bias inherent in single-model anomaly detection, we implement an ensemble voting strategy \cite{pacuit2011voting}. We employ a diverse set of unsupervised classifiers $\mathcal{M} = \{ \text{iForest}, \text{LOF}, \text{OC-SVM}, \text{EllipticEnvelope}, \dots \}$. Each classifier $M_j \in \mathcal{M}$ evaluates the SimHashed InfoPaths independently. The final classification of a path $P$ is determined by a majority vote:
\begin{equation}
V(P) = \mathbb{I}\left( \sum_{M_j \in \mathcal{M}} \mathbb{I}(M_j(P) = \text{anomalous}) > \frac{|\mathcal{M}|}{2} \right)
\end{equation}
Paths confirmed as anomalous by the majority are flagged for analyst review.

\subsubsection{InfoPath Retrieval and GPT-Powered Reporting}
\label{sec:infopath_retrieval}
The final step is to extract the most salient attack chains, termed \textit{InfoPaths}, from the graph. An InfoPath is defined as a directed sequence of causally linked process nodes $(v_1, v_2, \dots, v_k)$ representing a potential attack progression.

To retrieve the most critical paths, we adapt Dijkstra's shortest-path algorithm \cite{johnson1973note}. Since standard Dijkstra minimizes cost, and our refined weights $W_{mn}^{refined}$ represent *evidence* (where higher is more significant), we define the traversal cost function as the inverse of the evidence, further weighted by the community score to favor cohesive attack phases:
\begin{equation}
Cost(e_{mn}) = \frac{1}{W_{mn}^{refined} \times S_{C_{community(n)}}}
\end{equation}
By minimizing this cost, the algorithm effectively maximizes the cumulative evidence along the path, prioritizing InfoPaths with high structural rarity, centrality, known threat indicators, and community coherence.

To mitigate alert fatigue, the Top-$k$ retrieved InfoPaths are fed into a Retrieval-Augmented Generation (RAG) framework~\cite{peric2024cyberllmrag} powered by Llama-2.
The model augments the raw path data with context from external threat intelligence reports, generating a concise, prioritized snapshot alarm that explains the attack intent and recommended actions.

\section{Evaluation}
\label{sec:evaluation}
The evaluation is divided into \textbf{overall evaluation} and \textbf{ablation evaluation}. 
The \textbf{overall evaluation} compares our approach with state-of-the-art methods using precision, recall, and $F_1$-score on selected academic and industrial attack datasets.
The \textbf{ablation evaluation} conducts a detailed functional analysis to assess the effectiveness of each component in our solution, providing deeper insights into its impact and performance on discussed RQs and limitations.

\subsection{Attack Datasets}

%In this section, we begin our evaluation on \textsc{DefendCLI} using established academic benchmark datasets. Our goal is to showcase the effectiveness of our proposed method by presenting three key aspects: the results of the command-line differentiation analysis, the performance benchmarks, and the outcomes of report generation and recommendation.
%In addition, detailed results of report generation and recommendation and run-time performance are presented in Appendix \ref{app:runtime}, showcasing \textsc{DefendCLI}'s excellence on academic benchmark datasets.

\subsubsection{\textbf{DARPA Engagement Series Datasets (Linux and Unix-Like Servers)}}
%\circledwhite{1} \textbf{{\uline{E3 Datasets.}}}
The DARPA Transparent Computing Engagement Series (TCES) \cite{TCP} are designed to support cybersecurity research by providing real-world cyber attack traces, system logs, and network activity. 
Due to most of related work are evaluated on the Engagement-3 (E3) dataset, to make a fair comparison, we also use the E3 dataset as the benchmark. 
The E3 dataset captures multi-user cyber attack scenarios with a mix of legitimate and malicious activities, including privilege escalation, lateral movement, and data exfiltration, making it ideal for anomaly detection and forensic analysis. 
It includes host-level logs, file system events, system calls, and process activities, making them valuable for developing machine learning models, behavioral analytics, and proactive cyber defense strategies.
Based on the attack scenarios, the E3 dataset is divided into E3-Trace, E3-THEIA and E3-CADETS.

\begin{table}[t]
\caption{GitHub Availability, Reproducibility and Real-Time Detection of Related Work}
\centering
\begin{tabular}{@{}l|cccc@{}}
\toprule
\textbf{Method}     & \textbf{A} & \textbf{RP} & \textbf{RT} &\textbf{C} \\ \midrule
\textsc{Holmes} \cite{milajerdi2019holmes}     
& \textcolor[HTML]{FF0000}{\ding{55}}
& \textcolor[HTML]{008000}{\ding{51}}
& \textcolor[HTML]{008000}{\ding{51}}
& S\&P
\\
\midrule
\textsc{Unicorn} \cite{han2020unicorn}    
& \textcolor[HTML]{008000}{\ding{51}}
& \textcolor[HTML]{008000}{\ding{51}}
& \textcolor[HTML]{008000}{\ding{51}}
& NDSS
\\ 
\midrule
\textsc{DepComm} \cite{xu2022depcomm}   
& \textcolor[HTML]{008000}{\ding{51}}
& \textcolor[HTML]{008000}{\ding{51}}
& \textcolor[HTML]{FF0000}{\ding{55}}
& S\&P
\\ 
\midrule
\textsc{Prographer} \cite{yang2023prographer} 
& \textcolor[HTML]{FF0000}{\ding{55}}
& \textcolor[HTML]{FF0000}{\ding{55}}
& \textcolor[HTML]{FF0000}{\ding{55}}
& Usenix Security
\\
\midrule
\textsc{Nodlink} \cite{li2023nodlink}   
& \textcolor[HTML]{008000}{\ding{51}}
& \textcolor[HTML]{008000}{\ding{51}}
& \textcolor[HTML]{008000}{\ding{51}}
& NDSS
\\
\midrule
%\textsc{Kairos} \cite{cheng2023kairos}     
%& \textcolor[HTML]{008000}{\ding{51}}
%& \textcolor[HTML]{008000}{\ding{51}}
%& \textcolor[HTML]{FF0000}{\ding{55}}
%\\
%\midrule
\textsc{DISTDET} \cite{dong2023distdet}    
& \textcolor[HTML]{FF0000}{\ding{55}}
& \textcolor[HTML]{FF0000}{\ding{55}}
& \textcolor[HTML]{008000}{\ding{51}}
& Usenix Security
\\
\bottomrule
\end{tabular}
\label{tab:github}
\end{table}

\begin{table}[t]
\caption{Licensed Commercial EDR Solutions}
\centering
\begin{tabular}{@{}l|cccc@{}}
\toprule
\textbf{Tool}     & \textbf{AV} & \textbf{WS} & \textbf{CTI} & \textbf{RT} \\ \midrule
\textsc{Microsoft Defender} \cite{defender}     
& \textcolor[HTML]{008000}{\ding{51}}
& \textcolor[HTML]{008000}{\ding{51}}
& \textcolor[HTML]{008000}{\ding{51}}
& \textcolor[HTML]{008000}{\ding{51}}
\\
\midrule
\textsc{Symantec EDR} \cite{symantec}    
& \textcolor[HTML]{008000}{\ding{51}}
& \textcolor[HTML]{008000}{\ding{51}}
& \textcolor[HTML]{008000}{\ding{51}}
& \textcolor[HTML]{008000}{\ding{51}}
\\ 
\midrule
\textsc{Karspersky EDR} \cite{kaspersky,kasperskyCloud}   
& \textcolor[HTML]{008000}{\ding{51}}
& \textcolor[HTML]{008000}{\ding{51}}
& \textcolor[HTML]{008000}{\ding{51}}
& \textcolor[HTML]{008000}{\ding{51}}
\\
\bottomrule
\end{tabular}
\label{tab:edr}
\end{table}

%\begin{table}[t]
%\caption{Graph-Level Result (E3-Trace)}
%\centering
%\begin{tabular}{@{}l|ccc@{}}
%\toprule
%\textbf{Method}     & \textbf{Precision} & \textbf{Recall} &\textbf{$F_1$ Score}  \\ \midrule
%\textsc{Holmes} \cite{milajerdi2019holmes}     
%& 0.15
%& 1
%& 0.26
%\\
%\midrule
%\textsc{Unicorn} \cite{han2020unicorn}    
%& 0.28
%& 1
%& 0.44
%\\
%\midrule
%\textsc{Kairos} \cite{cheng2023kairos}     
%& 0.29
%& 1
%& 0.45
%\\
%\midrule
%\textsc{DepComm} \cite{xu2022depcomm}   
%& 0.43
%& 0.98
%& 0.60
%\\
%\midrule
%\textsc{Nodlink} \cite{li2023nodlink}   
%& 0.67
%& 1
%& 0.80
%\\
%\midrule
%\textsc{DefendCLI}
%& 0.81
%& 1
%& 0.90
%\\
%\bottomrule
%\end{tabular}
%\label{tab:graph_e3_trace}
%\end{table}
\begin{table*}[t]
\caption{Node-Level Results on DARPA E3 Datasets}
\centering
\begin{tabular}{@{}l|ccc|ccc|ccc@{}}
\toprule
\multirow{2}{*}{\textbf{Method}} & \multicolumn{3}{c|}{\textbf{E3-Trace}} & \multicolumn{3}{c|}{\textbf{E3-THEIA}} & \multicolumn{3}{c}{\textbf{E3-CADETS}} \\
\cmidrule(lr){2-4} \cmidrule(lr){5-7} \cmidrule(lr){8-10}
& \textbf{Precision} & \textbf{Recall} & \textbf{$F_1$ Score}  
& \textbf{Precision} & \textbf{Recall} & \textbf{$F_1$ Score}  
& \textbf{Precision} & \textbf{Recall} & \textbf{$F_1$ Score}  
\\ \midrule
\textsc{Holmes} \cite{milajerdi2019holmes}     
& $1.35 \times 10^{-3}$ & 0.74 & $2 \times 10^{-3}$
& $3.61 \times 10^{-3}$ & 0.98 & 0.01
& $2.84 \times 10^{-3}$ & 0.95 & 0.01
\\
\midrule
\textsc{Unicorn} \cite{han2020unicorn}    
& $3.20 \times 10^{-5}$ & N/A & N/A
& $1.86 \times 10^{-4}$ & N/A & N/A
& $1.25 \times 10^{-4}$ & N/A & N/A
\\
\midrule
\textsc{DepComm} \cite{xu2022depcomm}   
& $2.3 \times 10^{-3}$ & 1 & $4.6 \times 10^{-3}$
& $1.2 \times 10^{-2}$ & 1 & 0.02
& $1.2 \times 10^{-2}$ & 1 & 0.02
\\
\midrule
\textsc{Nodlink} \cite{li2023nodlink}   
& 0.25 & 0.98 & 0.40
& 0.23 & 1 & 0.37
& 0.14 & 1 & 0.25
\\
\midrule
\textsc{DefendCLI}
& 0.41 & 1 & 0.58
& 0.38 & 1 & 0.55
& 0.2 & 1 & 0.33
\\
\bottomrule
\end{tabular}
\label{tab:node_e3_all}
\end{table*}

\begin{table}[t]
\caption{Attack Scenario Checklist (E3)}
\centering
\begin{tabular}{@{}l|c|c@{}}
\toprule
\textbf{Dataset}   & \textbf{Attack Scenario} & \textbf{TTPs}  
\\
\midrule
\multirow{3}{*}{TRACE} 
& Firefox Exploit via Ads & Drive-by Execution  
\\
\cmidrule{2-3}
& Malicious Extension Drops Payload & DLL Injection  
\\
\cmidrule{2-3}
& Phishing Email Installs Backdoor & Credential Theft  
\\
\midrule
\multirow{2}{*}{THEIA}
& Firefox Exploit Injects Script & Remote Execution  
\\
\cmidrule{2-3}
& Fake Extension Steals Passwords & Keylogging Attack  
\\ 
\midrule
\multirow{3}{*}{CADETS}
& Web Server Exploited Remotely & Process Injection  
\\
\cmidrule{2-3}
& Fake Update Installs Malware & Persistence Attack  
\\
\cmidrule{2-3}
& Lateral Movement via Credentials & Privilege Escalation  
\\ 
\bottomrule
\end{tabular}
\label{tab:attack_scenarios_e3}
\end{table}

\begin{figure}[t]
    \centering
    \captionsetup[subfigure]{font=footnotesize, justification=centering} % 让子图 caption 变小，可用 footnotesize 或 scriptsize
    % 第一行
    \begin{subfigure}{0.45\linewidth}
        \centering
        \includegraphics[width=\linewidth]{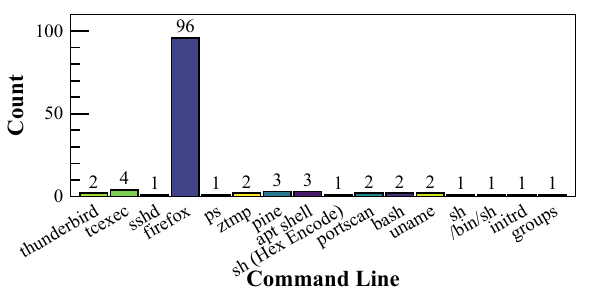}
        \caption{Attack-Related Command-Line (E3-TRACE)}
        \label{fig:cmd_count_trace}
    \end{subfigure}
    \hfill
    \begin{subfigure}{0.45\linewidth}
        \centering
        \includegraphics[width=\linewidth]{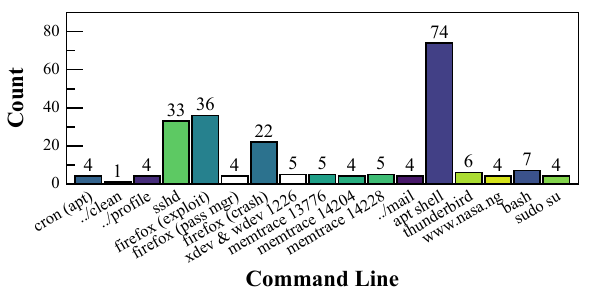}
        \caption{Attack-Related Command-Line (E3-THEIA)}
        \label{fig:cmd_count_theia}
    \end{subfigure}

    \vspace{10pt} % 调整行间距

    % 第二行
    \begin{subfigure}{0.45\linewidth}
        \centering
        \includegraphics[width=\linewidth]{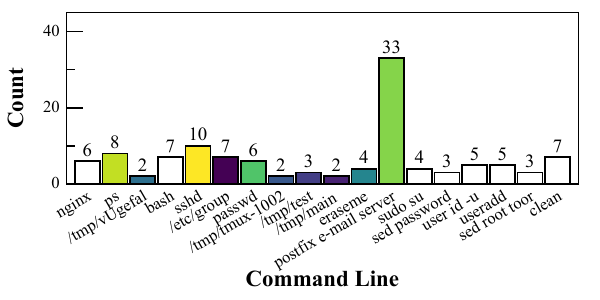}
        \caption{Attack-Related Command-Line (E3-CADETS)}
        \label{fig:cmd_count_cadets}
    \end{subfigure}
    \hfill
    \begin{subfigure}{0.45\linewidth}
        \centering
        \includegraphics[width=\linewidth]{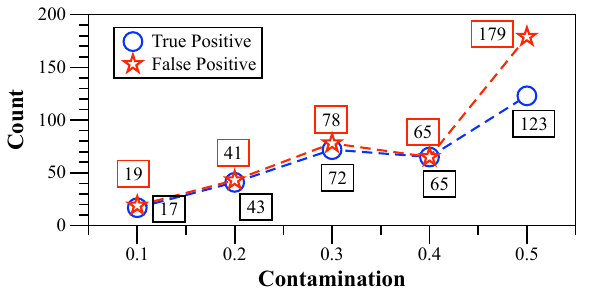}
        \caption{InfoPath-Level Result (E3-TRACE)}
        \label{fig:result_trace}
    \end{subfigure}

    \vspace{10pt} % 调整行间距

    % 第三行
    \begin{subfigure}{0.45\linewidth}
        \centering
        \includegraphics[width=\linewidth]{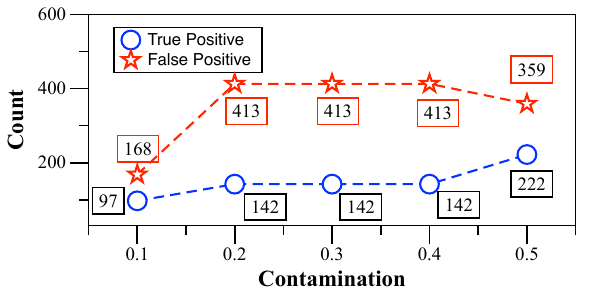}
        \caption{InfoPath-Level Result (E3-THEIA)}
        \label{fig:result_theia}
    \end{subfigure}
    \hfill
    \begin{subfigure}{0.45\linewidth}
        \centering
        \includegraphics[width=\linewidth]{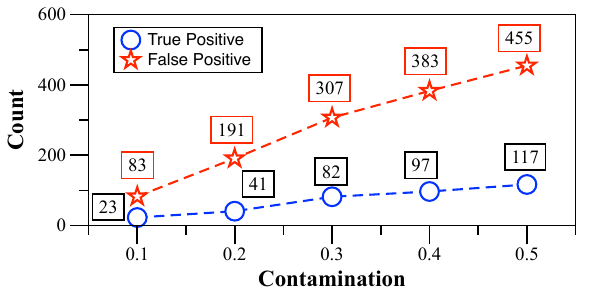}
        \caption{InfoPath-Level Result (E3-CADETS)}
        \label{fig:result_cadets}
    \end{subfigure}

    \caption{Comparison of attack-related command-line and InfoPath-Level results on E3 datasets.}
    \label{fig:overall_results}
\end{figure}

\subsubsection{\textbf{Industrial Real-Time Detection (Microsoft Windows End-User and Servers)}}
The Industrial Real-Time Detection dataset \cite{sangfor} is designed to support cybersecurity research by providing real-world, real-time telemetry from virtual machines (VMs) hosted on Microsoft Azure, enabling the evaluation of real-time detection and response capabilities for defense solutions. 
The dataset captures system logs, network activity, and security events in an industrial environment, simulating active cyber threats and operational challenges as they unfold. By including both benign and malicious activities, it allows researchers to study and enhance real-time threat detection, anomaly identification, and rapid intrusion response in cloud-based infrastructures. 
With realistic attack scenarios such as unauthorized access, privilege escalation, and lateral movement, this dataset is particularly valuable for advancing machine learning-driven security solutions, continuous monitoring systems, and proactive cyber defense strategies, emphasizing the importance of real-time detection and mitigation in industrial cloud environments.

\begin{table*}[t]
\caption{Attack-TTPs and Node-Level Results on Industrial Real-Time Detection}
\centering
\begin{tabular}{@{}l|ccc|ccc|ccc@{}}
\toprule
\multirow{2}{*}{\textbf{Method}} & \multicolumn{3}{c|}{\textbf{Attack Scenario \#1 (A1)}} & \multicolumn{3}{c|}{\textbf{Attack Scenario \#2 (A2)}} & \multicolumn{3}{c}{\textbf{Attack Scenario \#3 (A3)}} \\
\cmidrule(lr){2-4} \cmidrule(lr){5-7} \cmidrule(lr){8-10}
& \textbf{Precision} & \textbf{Recall} & \textbf{$F_1$ Score}  
& \textbf{Precision} & \textbf{Recall} & \textbf{$F_1$ Score}  
& \textbf{Precision} & \textbf{Recall} & \textbf{$F_1$ Score}  
\\ \midrule
\textsc{Microsoft Defender} \cite{defender}     
& 0.04 & 1 & 0.08
& 0 & 1 & 0
& 0 & 1 & 0
\\
\midrule
\textsc{Symantec EDR} \cite{symantec}    
& block & block & block
& 0.13 & 1 & 0.23
& 0.04 & 1 & 0.08
\\
\midrule
\textsc{Kasparsky EDR} \cite{kaspersky}   
& block & block & block
& block & block & blcok
& 0.09 & 1 & 0.10
\\
\midrule
\textsc{Nodlink \cite{li2023nodlink} (TTPs)}
& 1 & 1 & 1
& 1 & 1 & 1
& 1 & 1 & 1
\\
\midrule
\textsc{Nodlink \cite{li2023nodlink} (Node-Level)}
& 0.04 & 1 & 0.08
& 0.19 & 1 & 0.32
& 0.05 & 1 & 0.10
\\
\midrule
\textsc{DefendCLI (TTPs)}
& 1 & 1 & 1
& 1 & 1 & 1
& 1 & 1 & 1
\\
\midrule
\textsc{DefendCLI (Node-Level)}
& 0.08 & 1 & 0.15
& 0.38 & 1 & 0.55
& 0.18 & 1 & 0.31
\\
\bottomrule
\end{tabular}
\label{tab:node_real_time_all}
\end{table*}

\begin{table}[t]
\caption{Attack Scenario Checklist (Industrial)}
\centering
\begin{tabular}{@{}l|c|c@{}}
\toprule
\textbf{Scenario}   & \textbf{APT} & \textbf{TTPs}  
\\
\midrule
1   
& Silent Lynx \cite{sliver,Kazakhstan}
& PasteBin Multi-Staged Powershell
\\
\midrule
2
& DriftingCloud \cite{DriftingCloud,chopperwebshell}
& MITM PHP Chopper Webshell 
\\ 
\midrule
3
& Evilnum \cite{pythonrat}
& Spear Phishing Python RAT
\\ 
\bottomrule
\end{tabular}
\label{tab:payload}
\end{table}

\subsection{GitHub Availability, Reproducibility and Real-Time Detection}
Table \ref{tab:github} shows the \textbf{availability (A)}, \textbf{reproducibility (RP)}, and \textbf{real-time detection capability (RT)} of related work based on \textbf{big-4 security conferences (C)}.
The research work of \textsc{DISTDET} \cite{dong2023distdet} and \textsc{Prographer} \cite{yang2023prographer} are not publicly available on GitHub or other sources.
Therefore, our comparative study focuses on the remaining work. 
The verification of availability was conducted on 7 February 2025.

\subsection{Licensed Commercial EDR Solutions}
Table \ref{tab:edr} presents the detection capabilities of commercial EDR solutions across four key security features: \textbf{anti-virus (AV)}, \textbf{web security (WS)}, \textbf{cyber threat intelligence (CTI)}, and \textbf{real-time (RT)} protection. 
For our evaluation, we obtained commercial licenses for Microsoft Defender \cite{defender}, Symantec EDR \cite{symantec}, and Kaspersky EDR \cite{kaspersky,kasperskyCloud}. Each EDR solution was updated to the latest version, ensuring the most recent threat database was used during testing.

\subsection{Attack Provenance Evaluation Criterion}
We use a multi-level criterion to evaluate our work, which is introduced as follows:

\begin{itemize}[left=0pt]
    \item \textbf{InfoPath-Level:}
    The InfoPath-level evaluation criterion was introduced by \textsc{DefendCLI}, which detects anomalous attack chains by analyzing multiple paths between source and destination nodes. InfoPaths represent the interactions between system behaviors and serve as a core metric in both real-time industrial detection scenarios and our ablation studies.
    \item \textbf{Node-Level}
    The node-level criterion, first introduced in \textsc{Nodlink}~\cite{li2023nodlink}, addresses the limitations of prior approaches that evaluated entire attack provenance graphs often achieving perfect precision while inadvertently including many non-attack-related events. Instead, this criterion focuses on identifying and quantifying attack-relevant nodes within InfoPaths, offering a more granular and accurate reflection of actual attack activity.
    \item \textbf{TTPs-Level}
    The TTPs-level criterion is used to compare detection capabilities against commercial EDR tools. Since many EDRs terminate execution upon detecting an attack, limiting further behavior observation, we assess the number of detected activities associated with adversarial tactics, techniques, and procedures (TTPs). This allows for a direct performance comparison between our solution and commercial detection tools.
\end{itemize}

\subsection{Overall Evaluation}
\subsubsection{\textbf{Detection Results on DARPA E3 Benchmark}}
Table \ref{tab:attack_scenarios_e3} presents the threat model of the three datasets in the E3 benchmark, while Table \ref{tab:node_e3_all} compares node-level detection results with state-of-the-art research.
Our analysis shows that while all existing methods can detect E3 attack scenarios, their node-level results vary significantly. 
This discrepancy highlights the presence of substantial non-attack-related information in their alarms, leading to lower detection precision and reduced $F_1$ scores.

Our solution \textsc{DefendCLI}, improves detection by focusing on command-line-level analysis, achieving a precision improvement of 1.6 $\times$ over the best existing method \textsc{Nodlink}~\cite{li2023nodlink}. 
Figure \ref{fig:overall_results} presents the results at the InfoPath-level and the command-line activities related to the detected attack identified by \textsc{DefendCLI}.
This improvement demonstrates that \textsc{DefendCLI} provides more concise results, detecting E3 attack scenarios with fewer nodes while delivering more actionable insights for threat incident verification, significantly enhancing detection efficiency.

\begin{figure}[t]
    \centering
    \captionsetup[subfigure]{font=footnotesize, justification=centering} % 让子图 caption 变小，可用 footnotesize 或 scriptsize
    % 第一行
    \begin{subfigure}{0.45\linewidth}
        \centering
        \includegraphics[width=\linewidth]{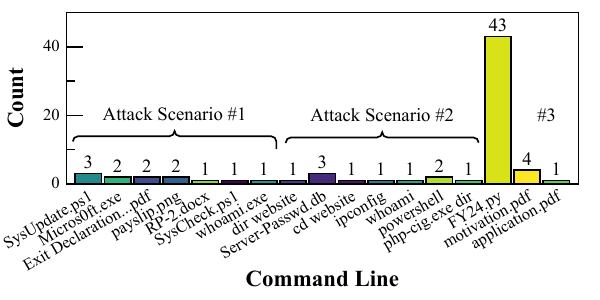}
        \caption{Attack-Related Command-Line (Real-Time)}
        \label{fig:cmd_real_time}
    \end{subfigure}
    \hfill
    \begin{subfigure}{0.45\linewidth}
        \centering
        \includegraphics[width=\linewidth]{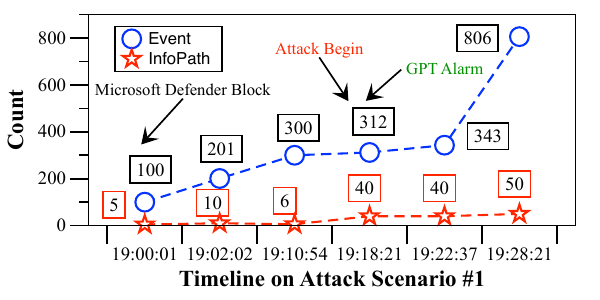}
        \caption{Real-Time InfoPath Alarm (Attack Scenario \#1)}
        \label{fig:result_a1}
    \end{subfigure}

    \vspace{10pt} % 调整行间距

    % 第二行
    \begin{subfigure}{0.45\linewidth}
        \centering
        \includegraphics[width=\linewidth]{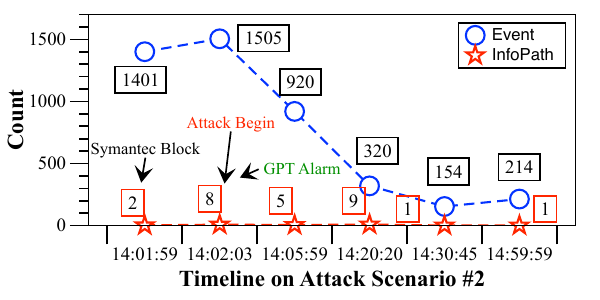}
        \caption{Real-Time InfoPath Alarm (Attack Scenario \#2)}
        \label{fig:result_a2}
    \end{subfigure}
    \hfill
    \begin{subfigure}{0.45\linewidth}
        \centering
        \includegraphics[width=\linewidth]{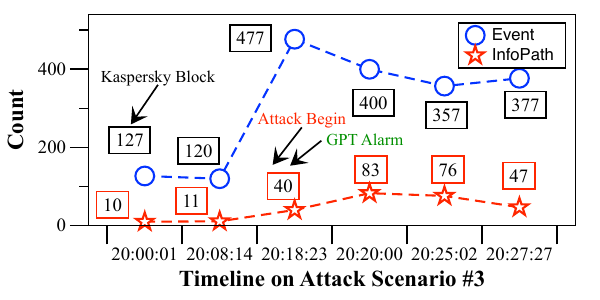}
        \caption{Real-Time InfoPath Alarm (Attack Scenario \#3)}
        \label{fig:result_trace}
    \end{subfigure}
    \caption{Comparison of attack-related command-line and InfoPath alarms in real-time detection.}
    \label{fig:real-time-infopath}
\end{figure}

\subsubsection{\textbf{Detection Results in the Industrial Real-Time Detection}}
Table \ref{tab:payload} presents the threat model for industrial real-time detection, incorporating modern APT threats relevant to 2025. 
These advanced threats are designed to evade modern defense strategies, enabling a more comprehensive evaluation of our solution’s detection capabilities in complex industrial environments.

Table \ref{tab:node_real_time_all} compares the node-level and TTP-level detection results of \textsc{DefendCLI} against modern commercial EDRs and the best existing method, \textsc{Nodlink}~\cite{li2023nodlink}. 
The results indicate that Microsoft Defender is bypassed in all three attack scenarios, Symantec EDR fails to detect attack scenarios 2 and 3, and Kaspersky EDR is bypassed by attack scenario 3.
At the TTP level, both \textsc{DefendCLI} and \textsc{Nodlink} successfully detect all malicious activities, demonstrating the effectiveness of attack provenance-based solutions.
However, at the node level, \textsc{DefendCLI} achieves a 2.3$\times$ precision improvement over \textsc{Nodlink}, further highlighting its superior detection efficiency and performability.

Figure \ref{fig:real-time-infopath} illustrates the attack-related command-line activities and the number of real-time InfoPath alarms identified by \textsc{DefendCLI}.
The industrial environment presents greater complexity than the E3 datasets due to a significant number of active users generating normal activities during evaluation.
Despite this, our results show that \textsc{DefendCLI} maintains a sound alarm rate relative to the volume of generated events, demonstrating its practicality for real-world deployment.

\begin{figure}[t]
    \centering
    \captionsetup[subfigure]{font=footnotesize, justification=centering} 

    % First row
    \begin{subfigure}{0.45\linewidth}
        \centering
        \includegraphics[width=\linewidth]{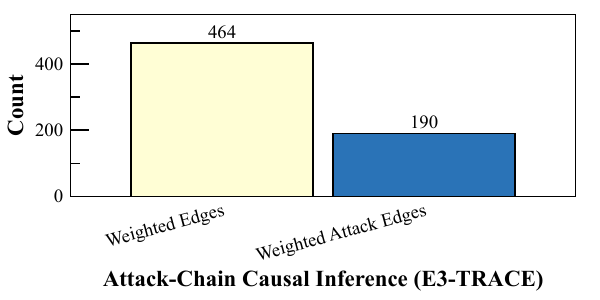}
        \caption{Attack-Chain Causal Inference (E3-TRACE)}
        \label{fig:acc-trace}
    \end{subfigure}
    \hfill
    \begin{subfigure}{0.45\linewidth}
        \centering
        \includegraphics[width=\linewidth]{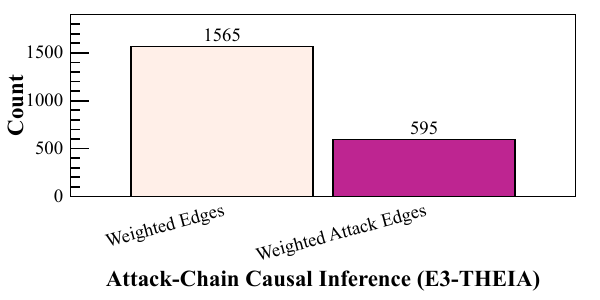}
        \caption{Attack-Chain Causal Inference (E3-THEIA)}
        \label{fig:acc-theia}
    \end{subfigure}

    \vspace{10pt} % Adjust row spacing

    % Second row
    \begin{subfigure}{0.45\linewidth}
        \centering
        \includegraphics[width=\linewidth]{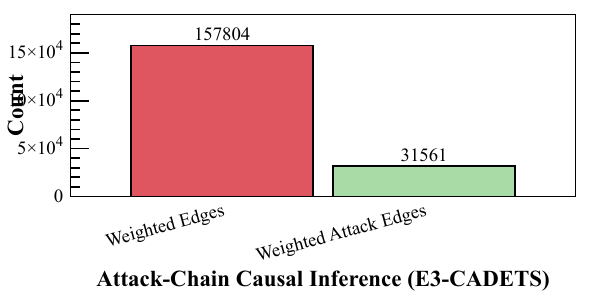}
        \caption{Attack-Chain Causal Inference (E3-CADETS)}
        \label{fig:acc-cadets}
    \end{subfigure}
    \hfill
    \begin{subfigure}{0.45\linewidth}
        \centering
        \includegraphics[width=\linewidth]{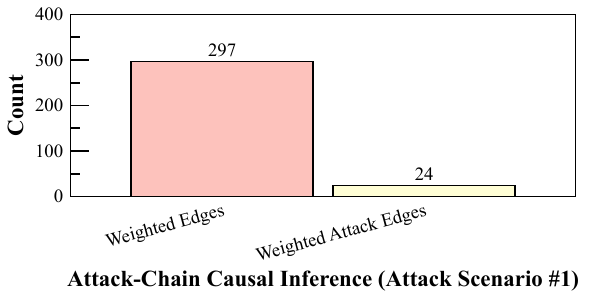}
        \caption{Attack-Chain Causal Inference (Attack Scenario \#1)}
        \label{fig:acc-1}
    \end{subfigure}

    \vspace{10pt} % Adjust row spacing

    % Third row
    \begin{subfigure}{0.45\linewidth}
        \centering
        \includegraphics[width=\linewidth]{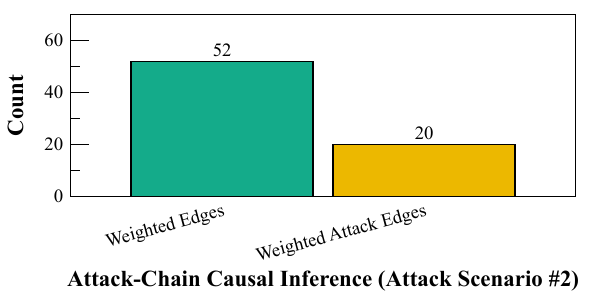}
        \caption{Attack-Chain Causal Inference (Attack Scenario \#2)}
        \label{fig:acc-2}
    \end{subfigure}
    \hfill
    \begin{subfigure}{0.45\linewidth}
        \centering
        \includegraphics[width=\linewidth]{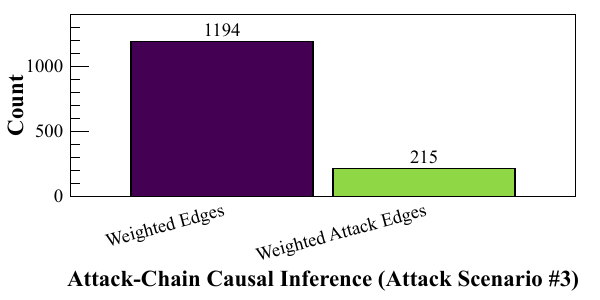}
        \caption{Attack-Chain Causal Inference (Attack Scenario \#3)}
        \label{fig:acc-3}
    \end{subfigure}

    \caption{Comparison of attack-chain causal inference based on the number of weighted edge count versus the number of attack-related weighted edge count.}
    \label{fig:acc_results}
\end{figure}

\subsection{Ablation Evaluation}

\subsubsection{\textbf{Evaluating the Interoperability of Attack Provenance Examination Through Attack-Clause Sketch and Evidence Awareness}}
Our solution incorporates a variety of sophisticated algorithms to enhance the detection capabilities of attack provenance analysis. 
In this section, we present a detailed assessment, breaking down the contribution of each module to demonstrate their individual effectiveness within our framework.

\textbf{$\blacksquare$ How well does attack-chain causal inference identify unusual elements within the attack provenance graph?}
The effectiveness of attack-chain causal inference in detecting unusual elements within the attack provenance graph is evaluated based on the comparison between the total number of weighted edges and the subset of edges directly associated with attack activities. Figure \ref{fig:acc_results} illustrates this comparison in different scenarios, highlighting the ability of our method to isolate suspicious elements from normal system interactions.

The results demonstrate that our causal inference model successfully assigns precise and meaningful weights to attack-related edges, ensuring that critical attack components are effectively highlighted while enabling efficient path traversal.
This mechanism allows for a clear distinction between malicious activities and benign system interactions, minimizing noise, and improving detection accuracy.
The observed disparity between the total number of weighted edges and the subset associated with attacks further validates the model’s ability to prioritize attack-relevant edges, effectively filtering out irrelevant connections and improving the interpretability of the attack provenance graph.

\begin{table}[t]
\caption{The Use and Not Use of Community Detection and SimHash Recommender on Node-Level $F_1$ Score}
\centering
\begin{tabular}{@{}l|cc|ccc@{}}
\toprule
\textbf{Dataset} & \multicolumn{2}{c|}{\textbf{Community Detection}} & \multicolumn{3}{c}{\textbf{SimHash \textcolor{blue}{(Blue)}}} \\
\cmidrule(lr){2-3} \cmidrule(lr){4-6}
                 & Y     & N     & W2V     & D2V     & FT    \\
\midrule
E3-TRACE       & \textcolor{blue}{0.58}  & 0.47  & \textcolor{blue}{0.58}  & 0.48  & 0.50   \\
\midrule
E3-THEIA       & \textcolor{blue}{0.55}  & 0.54  & 0.31  & 0.42  & \textcolor{blue}{0.55}   \\
\midrule
E3-CADETS      & \textcolor{blue}{0.33}  & 0.30  & \textcolor{blue}{0.33} & 0.31  & 0.32   \\
\midrule
A1             & \textcolor{blue}{0.15}  & 0.13  & 0.14   & 0.14  & \textcolor{blue}{0.15}   \\
\midrule
A2             & \textcolor{blue}{0.55}  & 0.31  & 0.48  & 0.30  & \textcolor{blue}{0.55}     \\
\midrule
A3             & 0.31  & 0.31  & \textcolor{blue}{0.31}  & 0.18  & 0.26    \\
\bottomrule
\end{tabular}
\label{tab:cd-sh}
\end{table}

\begin{figure}[t]
    \centering
    \captionsetup[subfigure]{font=footnotesize, justification=centering} % 让子图 caption 变小，可用 footnotesize 或 scriptsize
    % 第一行
    \begin{subfigure}{0.45\linewidth}
        \centering
        \includegraphics[width=\linewidth]{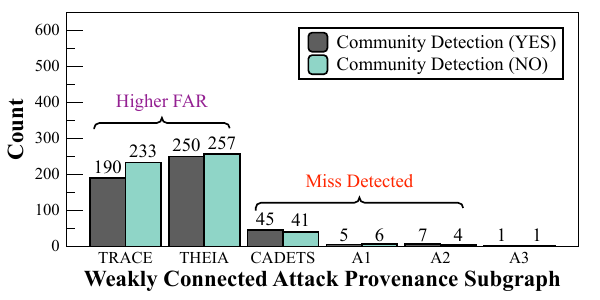}
        \caption{Community Detection}
        \label{fig:community-detection-c}
    \end{subfigure}
    \hfill
    \begin{subfigure}{0.45\linewidth}
        \centering
        \includegraphics[width=\linewidth]{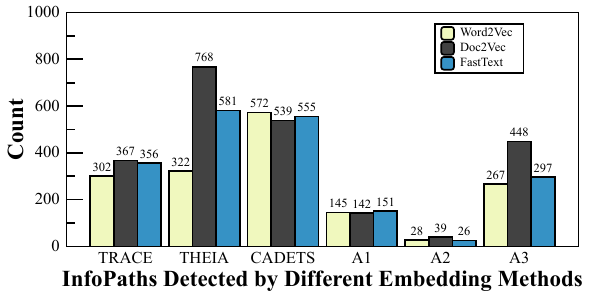}
        \caption{SimHash Recommender}
        \label{fig:simhash-recommender-c}
    \end{subfigure}
    \caption{Comparison of the use with or without community detection and SimHash recommender based on the count of InfoPaths detected.}
\end{figure}

\textbf{$\blacksquare$ How successfully does interphase attack association enhance the correlation of multi-staged attacks?}
Figure \ref{fig:community-detection-c} compares the use of community detection in analyzing the provenance of the attack, where Y/N denotes whether community detection is applied.
The \textcolor{blue}{blue} result highlights the better outcome.
We reconstructed the detected InfoPaths as weakly connected attack provenance subgraphs to assess how many reachable elements could be accurately correlated.

For the E3-TRACE and E3-THEIA datasets, we observed that without community detection, our solution results in a higher false alarm rate (FAR) while attempting to identify all attack activities. In contrast, for the E3-CADETS dataset, particularly in industrial attack scenarios \#1 and \#2, we found that while the absence of community detection may lead to fewer results, some attack activities go undetected.
However, for industrial attack scenario \#3, we observed no significant difference, as all elements were reachable regardless of the use of community detection.

Table \ref{tab:cd-sh} shows the node-level $F_1$ score with or without the use of community detection.
The experimental results can support our claim that the application of community detection improves correlation accuracy for identifying interphase attacks, leading to more precise and effective threat analysis.

\textbf{$\blacksquare$ How proficient is the SimHash recommender in distinguishing unknown command-line activities?}
Figure \ref{fig:simhash-recommender-c} compares the use of the SimHash recommender in analyzing the provenance of the attack, where W2V, D2V, and FT represent the use of Word2Vec, Doc2Vec, and FastText, respectively. The \textcolor{blue}{blue} result highlights the selection of the SimHash recommender.
For the E3-TRACE, E3-CADETS and attack scenario \#3, SimHash recommender selected Word2Vec as the method of vectorization because it has the maximum score to distinguish the command-line differentiation.
On the other hand, for the E3-THEIA, attack scenario \#1 and \#2, SimHash recommender selected FastText as the method of vectorization based on the same standard.

Similarly, we observed that for E3-TRACE and attack scenarios \#2 and \#3, SimHash generated fewer results while still detecting all attack behaviors, effectively reducing the FAR.  
In contrast, for E3-THEIA, E3-CADETS, and attack scenario \#1, SimHash produced a larger set of results, ensuring comprehensive attack detection.
However, other methods, despite yielding fewer results, failed to identify several malicious behaviors.

Table \ref{tab:cd-sh} presents the node-level $F_1$ score with and without the use of SimHash recommender.  
The results demonstrate that incorporating the SimHash recommender enhances context awareness by effectively differentiating command-line activities, leading to more precise threat analysis.

\begin{figure}[t]
       % \vspace{10pt} % 调整行间距
        
         \begin{subfigure}{0.45\linewidth}
        \centering
        \includegraphics[width=\linewidth]{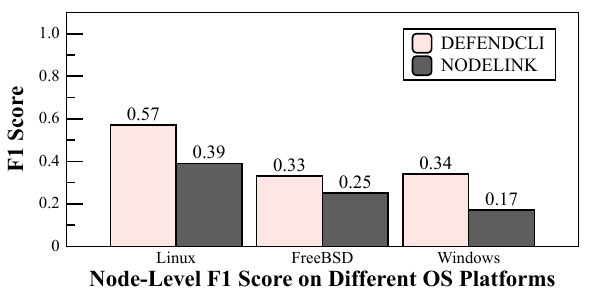}
        \caption{Diverse OS Platforms}
        \label{fig:os-platforms}
    \end{subfigure}
    \hfill
    \begin{subfigure}{0.45\linewidth}
        \centering
        \includegraphics[width=\linewidth]{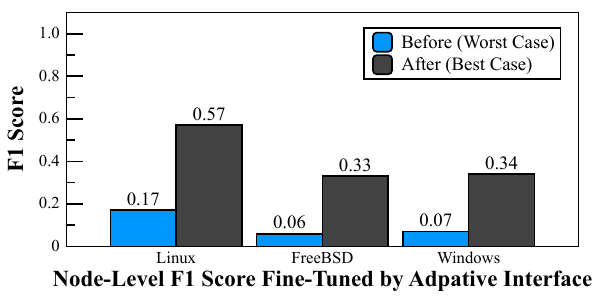}
        \caption{Adaptive Interface}
        \label{fig:adaptive}
    \end{subfigure}
    \caption{Comparison of the $F_1$ scores on diverse OS platforms and the use of adaptive interface.}

\end{figure}

\subsubsection{\textbf{Cross-Platform Detection Reliability: Performance Across Heterogeneous OS Environments}}
Figure \ref{fig:os-platforms} shows the node-level $F_1$ score across different OS environments.
In our evaluation, E3-TRACE and E3-THEIA were conducted on a Linux system, E3-CADETS was based on a FreeBSD system, and attack scenarios \#1 to \#3 were evaluated on a Windows system.
It can be seen that our solution benefited from the OS-specific detection mechanism, achieving superior detection performance compared to \textsc{Nodlink}, particularly on Linux, where system activities are highly dependent on command-line interactions. 
Unlike \textsc{Nodlink}, which employs a generic detection approach, our method adapts to system-specific behaviors by leveraging known attack signatures across different OS platforms.
This tailored strategy enhances detection accuracy and reliability, ensuring more precise identification of malicious activities.

\subsubsection{\textbf{Adaptive Detection in Dynamic Cyber Environments: Minimizing False Positives}}
Figure \ref{fig:adaptive} shows the node-level $F_1$ score before and after the use of the adaptive interface across different OS platforms.
The results of the worst-case and best-case scenarios are derived from a series of fine-tuning attempts conducted during our evaluation.

It can be observed that, AI detectors are highly sensitive to varying operational environments, necessitating dynamic fine-tuning to adapt effectively to changing conditions.
Similarly, each OS environment requires fine-tuning to achieve optimal detection results.

However, most research efforts \cite{li2023nodlink,han2020unicorn,xu2022depcomm,dong2023distdet,yang2023prographer} fail to provide a user-friendly, adaptive interface for fine-tuning their models. Instead, they rely on pre-defined parameters within their detection mechanisms, resulting in suboptimal performance in real-world detection scenarios.
Our adaptive interface effectively addresses this challenge, ensuring that our solution can be seamlessly applied in real-world detection scenarios.

\begin{figure}[t]
    \centering
    \captionsetup[subfigure]{font=footnotesize, justification=centering} 

    % First row
    \begin{subfigure}{0.45\linewidth}
        \centering
        \includegraphics[width=\linewidth]{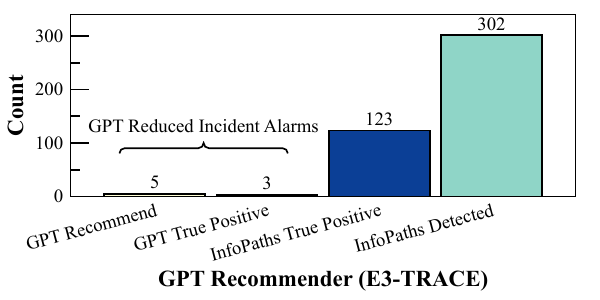}
        \caption{GPT Recommender Results on E3-TRACE}
        \label{fig:gpt_trace}
    \end{subfigure}
    \hfill
    \begin{subfigure}{0.45\linewidth}
        \centering
        \includegraphics[width=\linewidth]{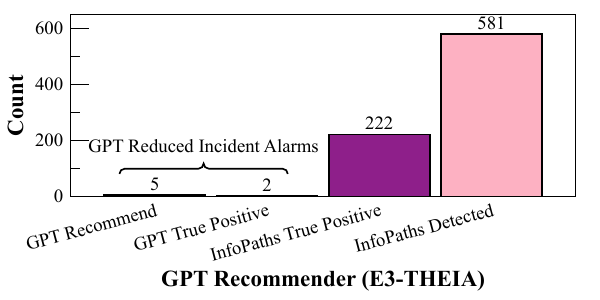}
        \caption{GPT Recommender Results on E3-THEIA}
        \label{fig:gpt_theia}
    \end{subfigure}

    \vspace{10pt} % Adjust row spacing

    % Second row
    \begin{subfigure}{0.45\linewidth}
        \centering
        \includegraphics[width=\linewidth]{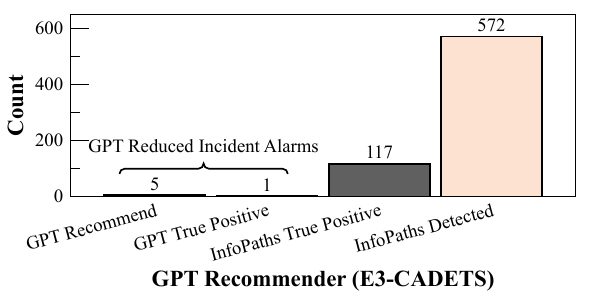}
        \caption{GPT Recommender Results on E3-CADETS}
        \label{fig:gpt_cadets}
    \end{subfigure}
    \hfill
    \begin{subfigure}{0.45\linewidth}
        \centering
        \includegraphics[width=\linewidth]{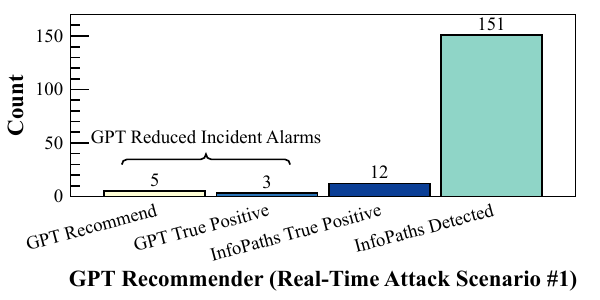}
        \caption{GPT Recommender Results on Real-Time Attack Scenario \#1}
        \label{fig:gpt_realtime1}
    \end{subfigure}

    \vspace{10pt} % Adjust row spacing

    % Third row
    \begin{subfigure}{0.45\linewidth}
        \centering
        \includegraphics[width=\linewidth]{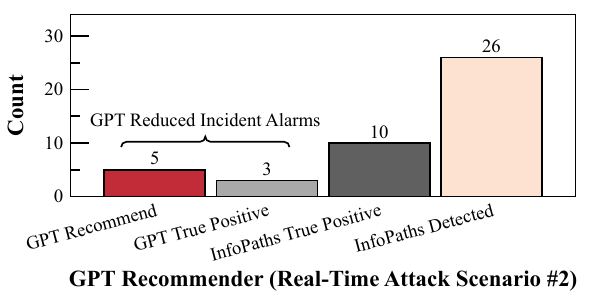}
        \caption{GPT Recommender Results on Real-Time Attack Scenario \#2}
        \label{fig:gpt_realtime2}
    \end{subfigure}
    \hfill
    \begin{subfigure}{0.45\linewidth}
        \centering
        \includegraphics[width=\linewidth]{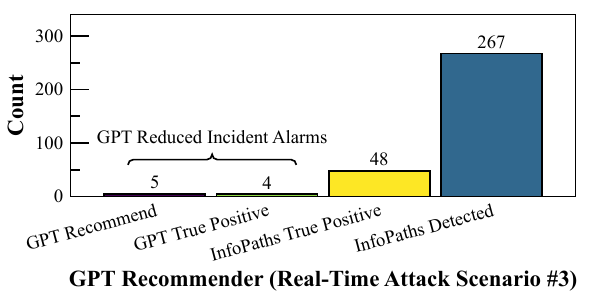}
        \caption{GPT Recommender Results on Real-Time Attack Scenario \#3}
        \label{fig:gpt_realtime3}
    \end{subfigure}

    \caption{Comparison of GPT recommender results across E3 datasets and real-time attack scenarios.}
    \label{fig:gpt_results}
\end{figure}

\begin{figure*}[t]
       % \vspace{10pt} % 调整行间距
        
         \begin{subfigure}{0.5\linewidth}
        \centering
        \includegraphics[width=\linewidth]{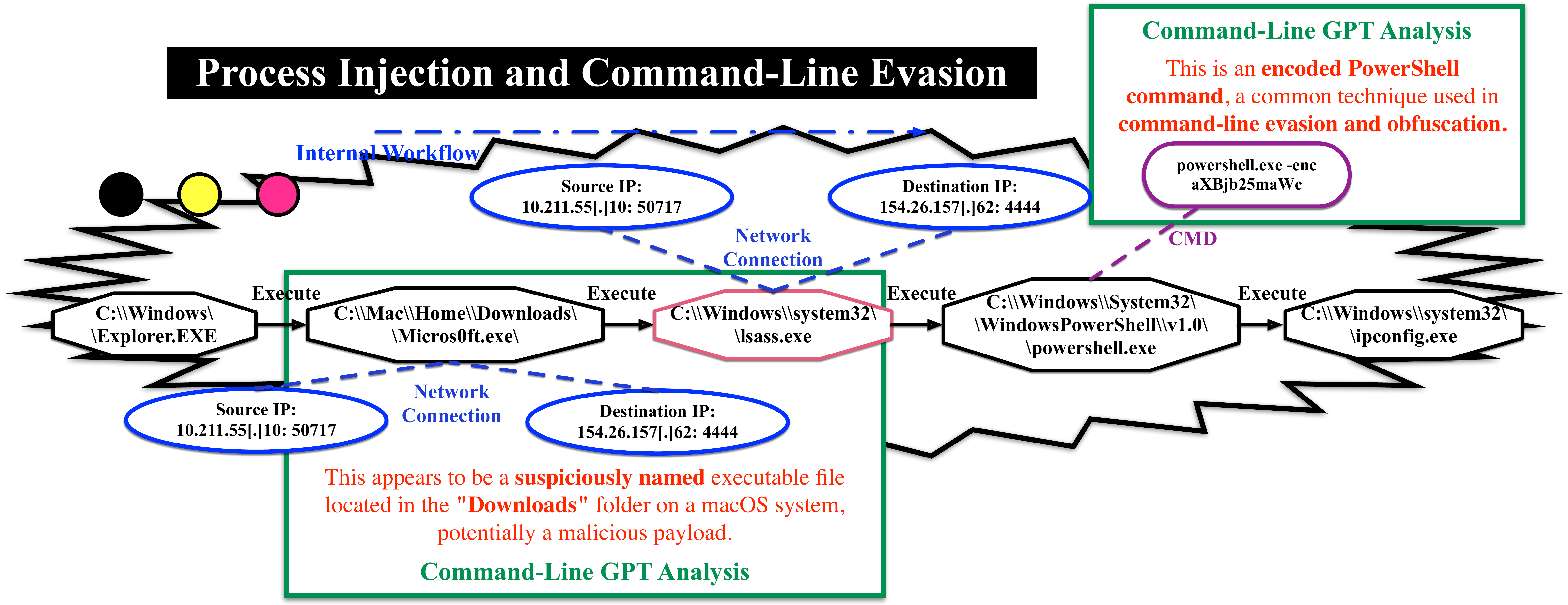}
        \caption{Progress Injection and Command-Line Evasion}
        \label{fig:v1}
    \end{subfigure}
    \hfill
    \begin{subfigure}{0.5\linewidth}
        \centering
        \includegraphics[width=\linewidth]{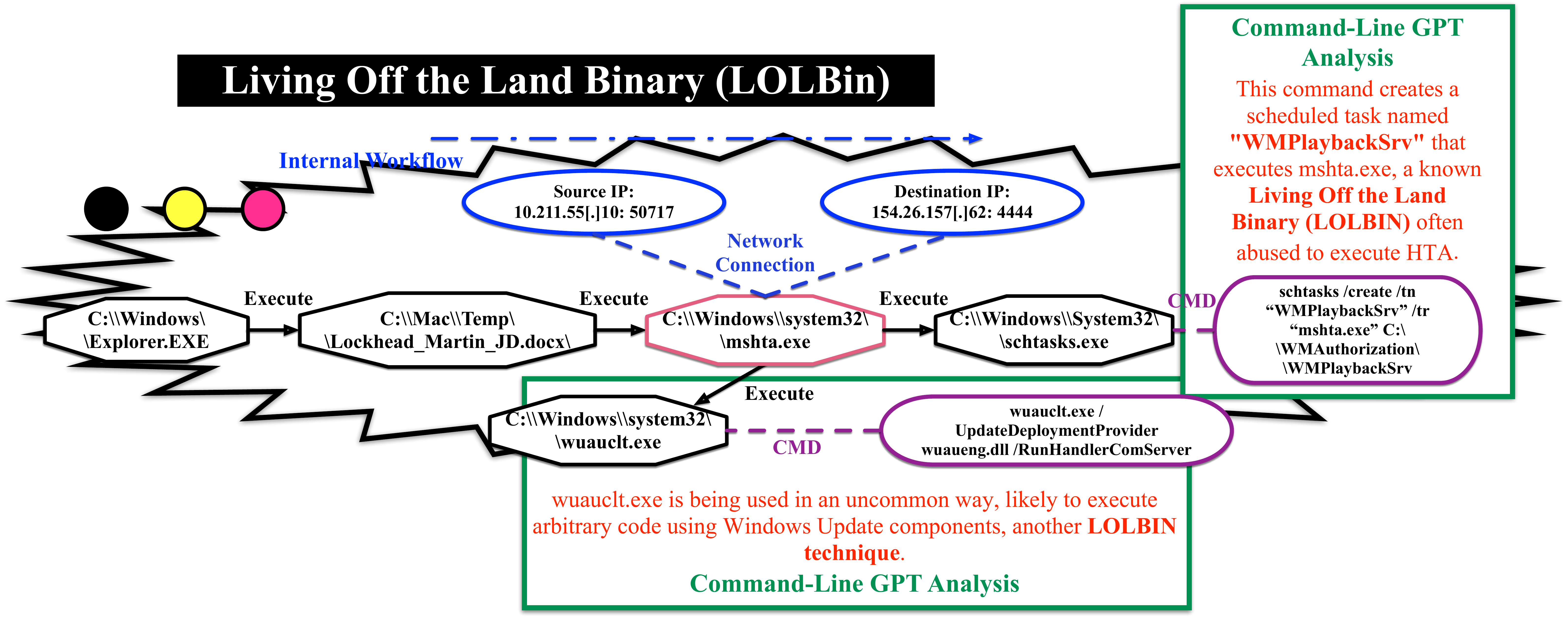}
        \caption{Living off the Land Binary (LOLBin)}
        \label{fig:v2}
    \end{subfigure}
    \caption{Visualization of (a) Process Injection and Command-Line Evasion (Silent Lynx \cite{sliver,Kazakhstan}); (b) Living off the Land Binary (LOLBin Evilnum \cite{pythonrat}); with GPT recommender.}
    \label{fig:v12}
\end{figure*}

\begin{figure*}[t]
       % \vspace{10pt} % 调整行间距
        
         \begin{subfigure}{0.3\linewidth}
        \centering
        \includegraphics[width=\linewidth]{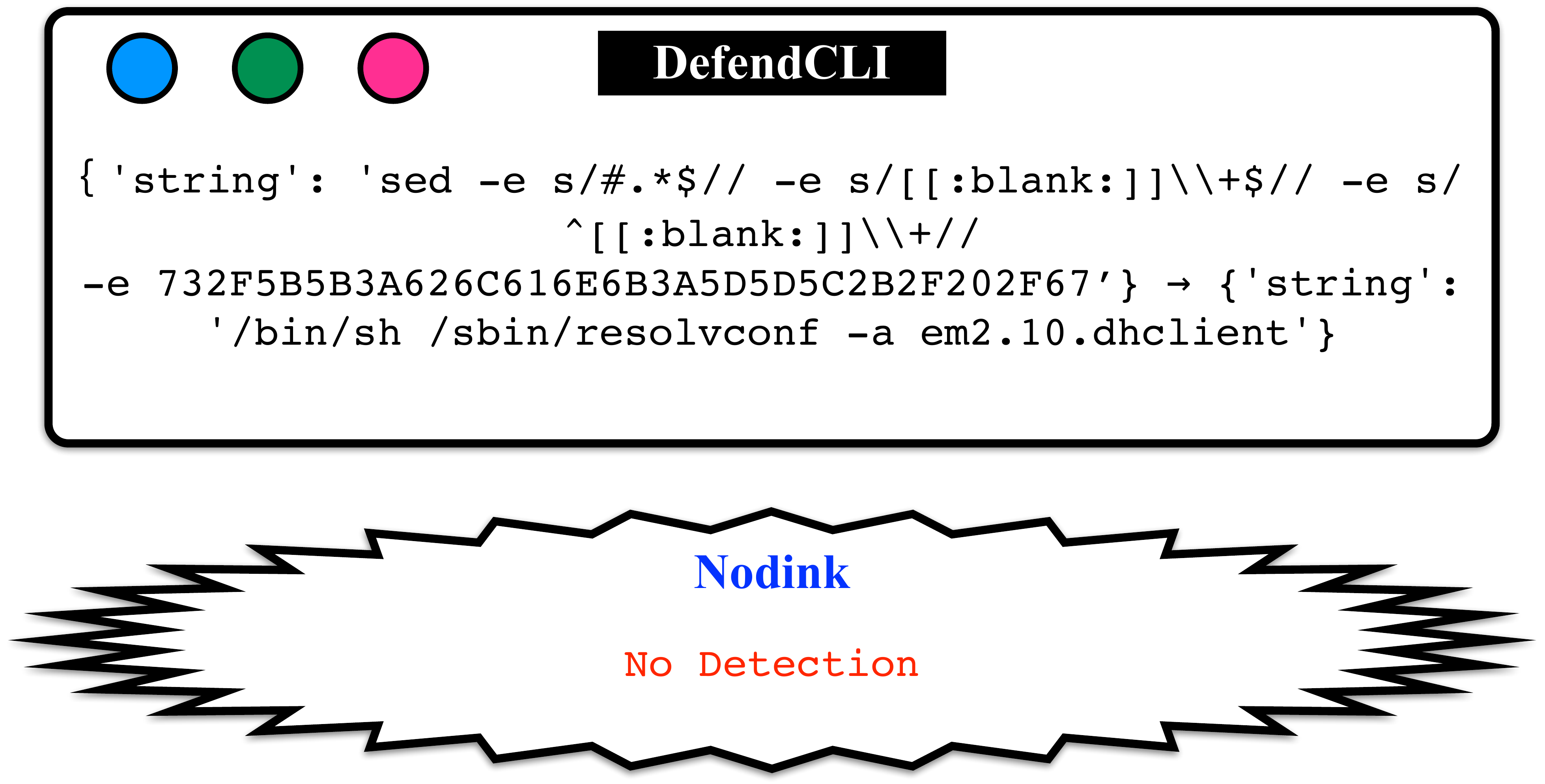}
        \caption{Advanced Command Transformation}
        \label{fig:transformation}
    \end{subfigure}
    \hfill
    \begin{subfigure}{0.3\linewidth}
        \centering
        \includegraphics[width=\linewidth]{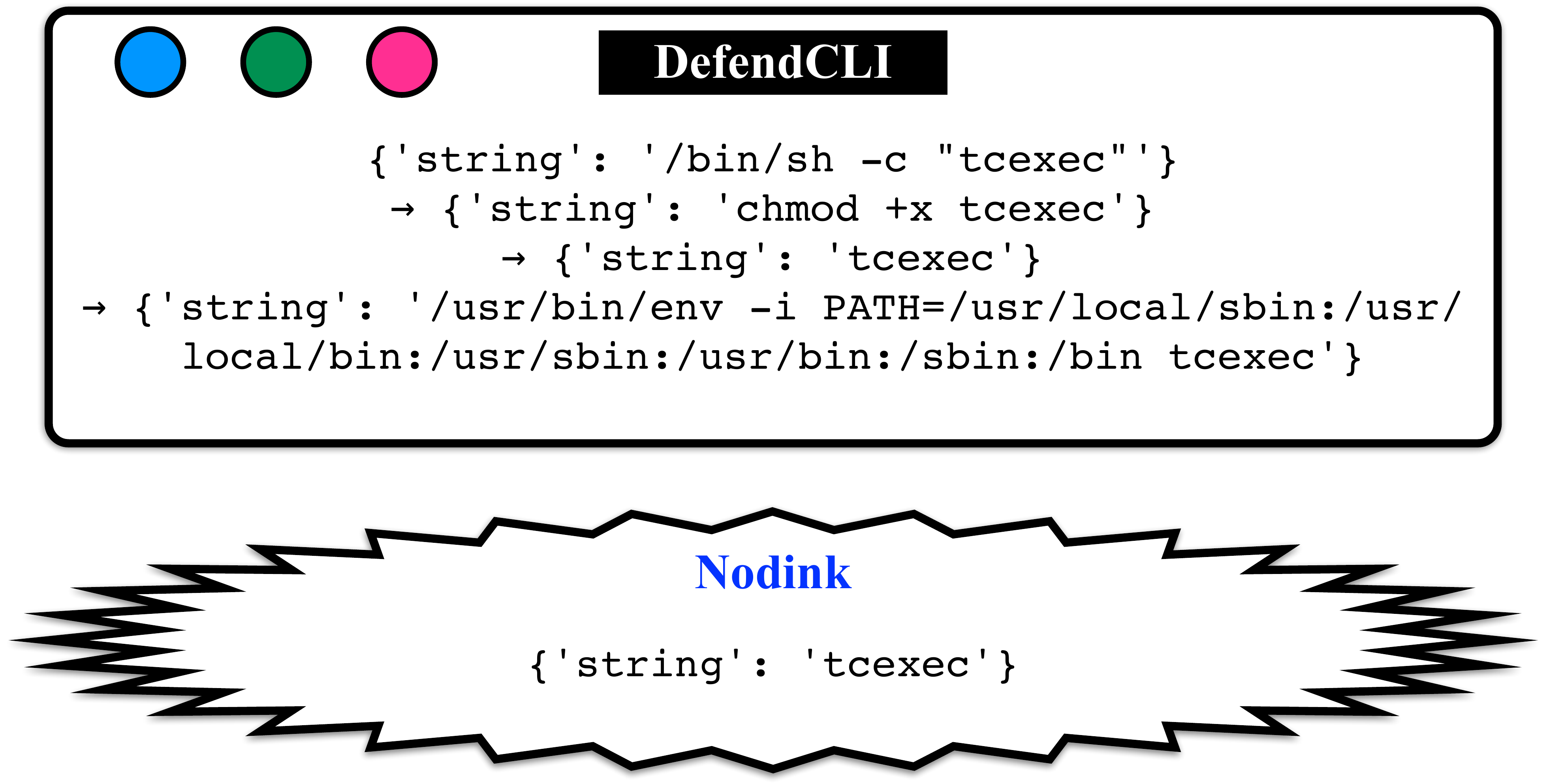}
        \caption{Backdoor Execution}
        \label{fig:tcexec}
    \end{subfigure}
    \hfill
    \begin{subfigure}{0.3\linewidth}
        \centering
        \includegraphics[width=\linewidth]{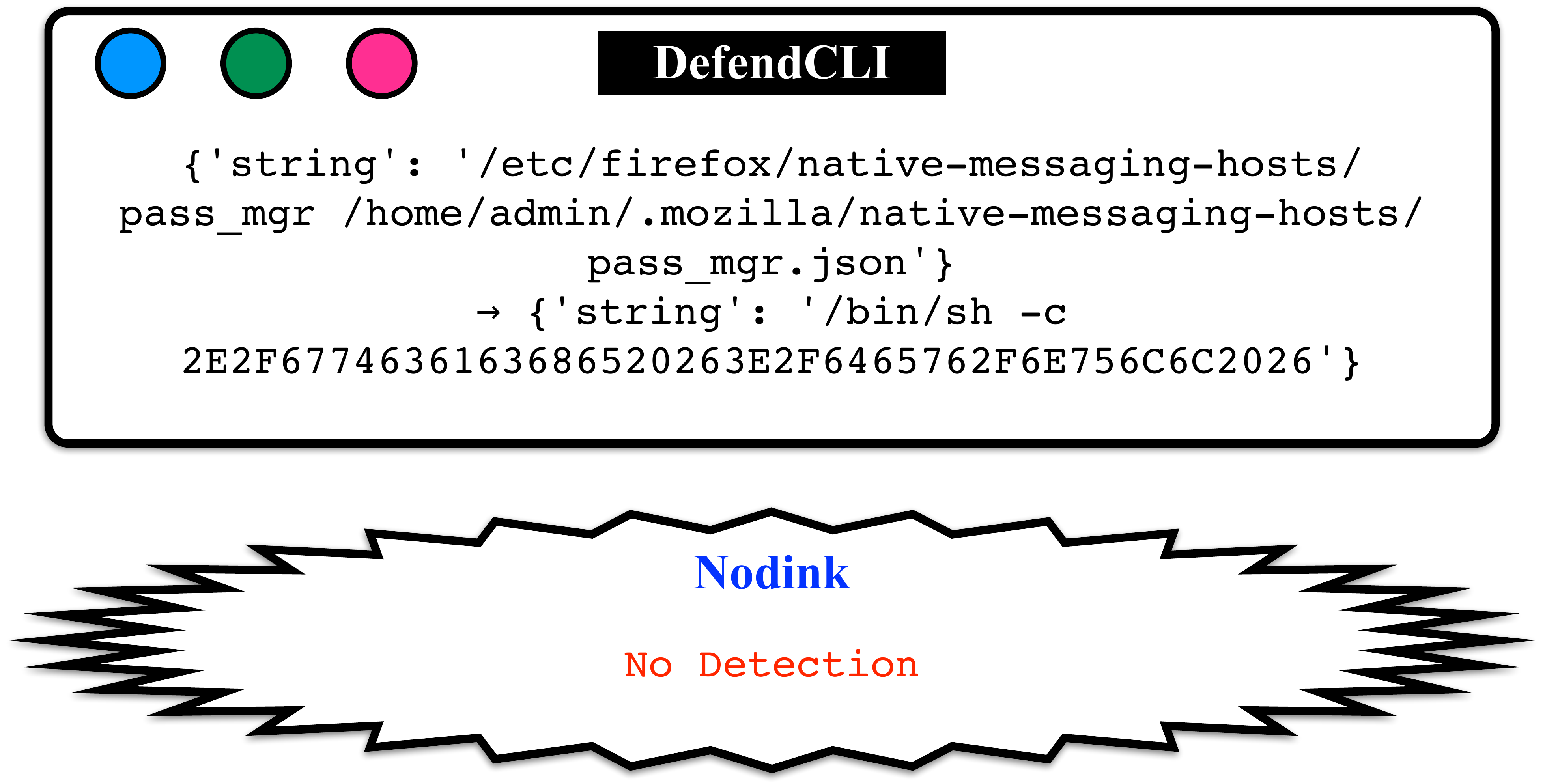}
        \caption{Base64 Shell Command Obfuscation}
        \label{fig:obfuscation}
    \end{subfigure}

      \vspace{10pt}
      
         \begin{subfigure}{0.3\linewidth}
        \centering
        \includegraphics[width=\linewidth]{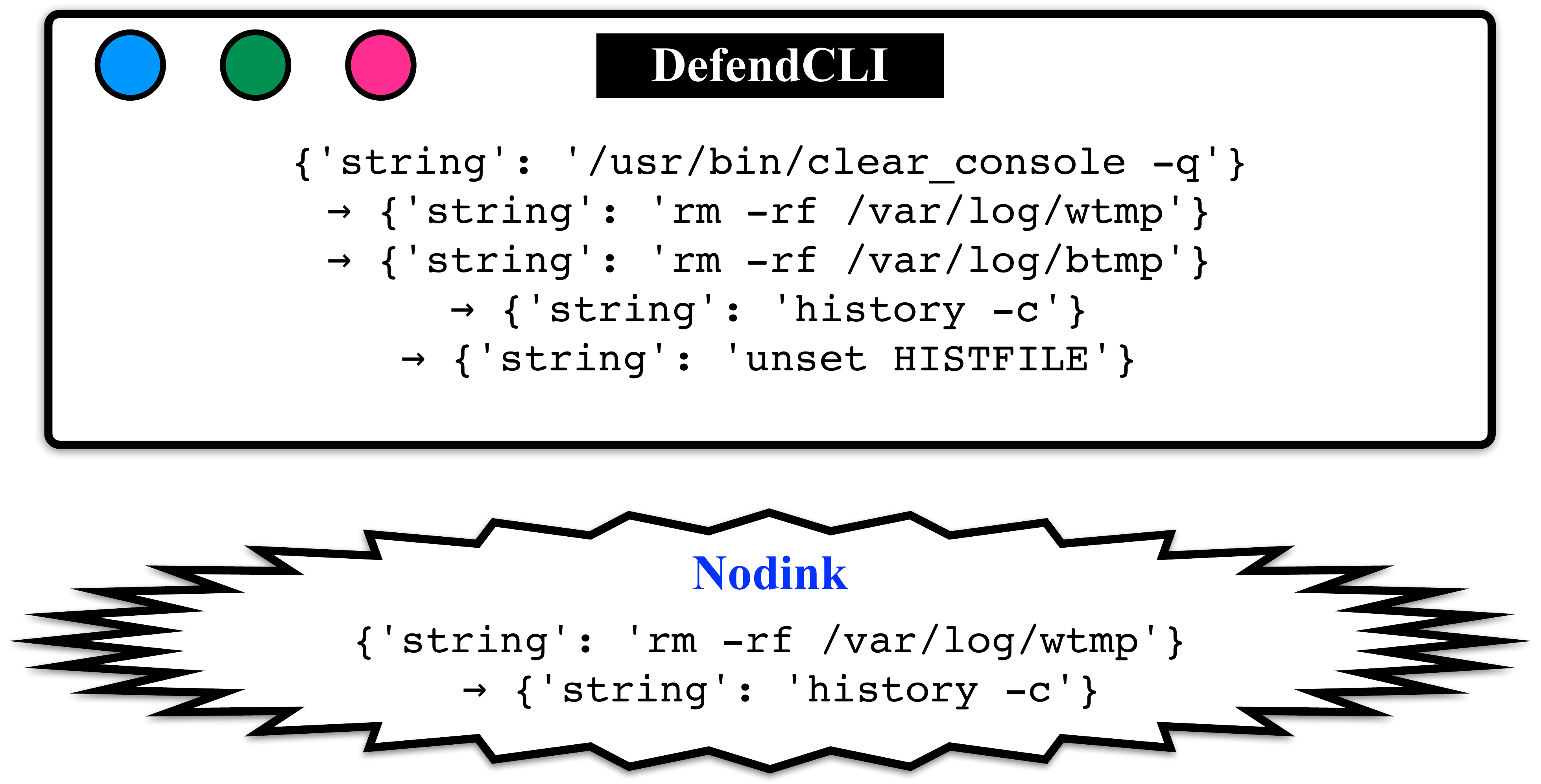}
        \caption{Log Deletion}
        \label{fig:log_deletion}
    \end{subfigure}
    \hfill
    \begin{subfigure}{0.3\linewidth}
        \centering
        \includegraphics[width=\linewidth]{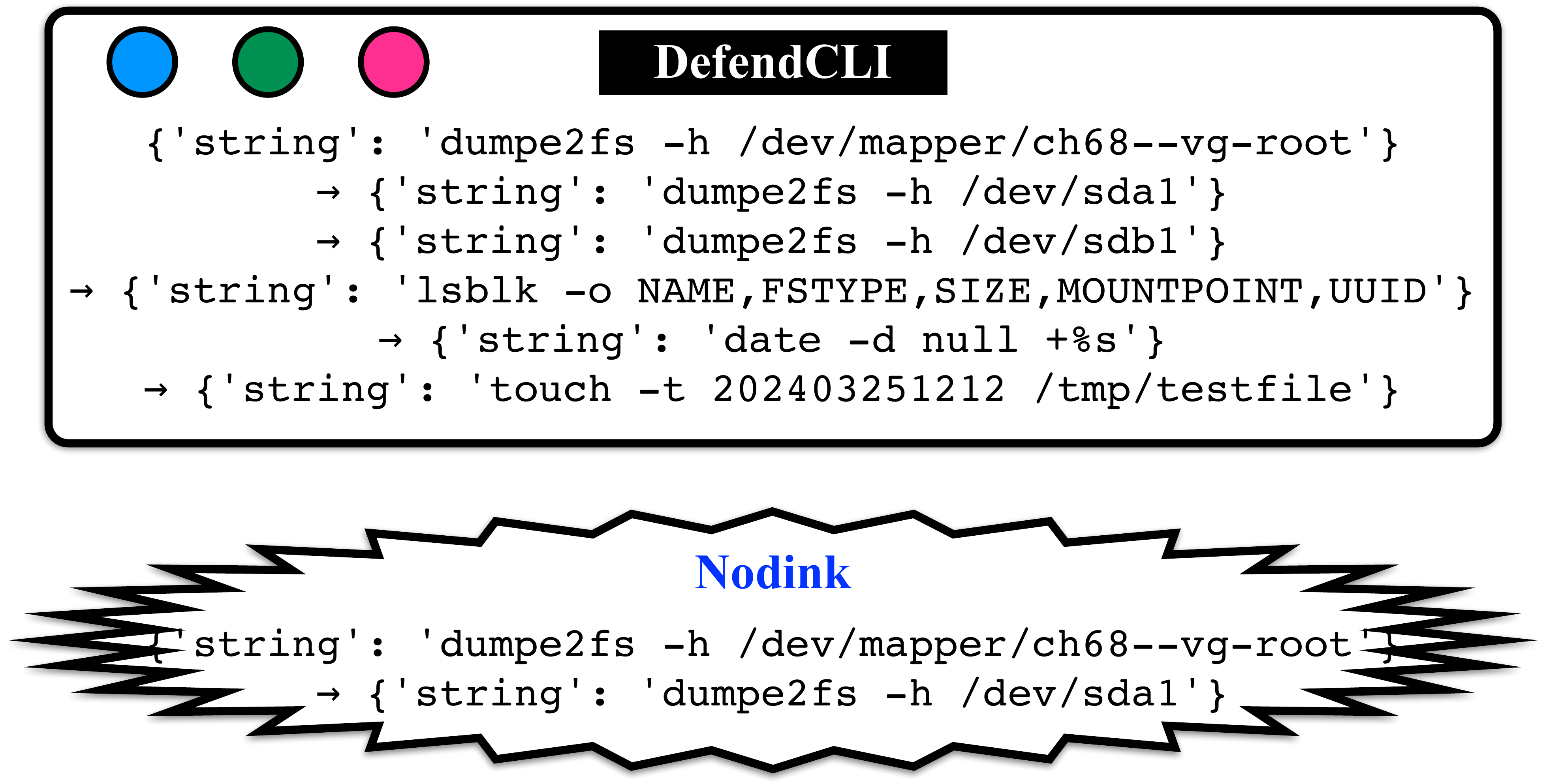}
        \caption{Timestamp Forgery}
        \label{fig:time_forgery}
    \end{subfigure}
    \hfill
    \begin{subfigure}{0.3\linewidth}
        \centering
        \includegraphics[width=\linewidth]{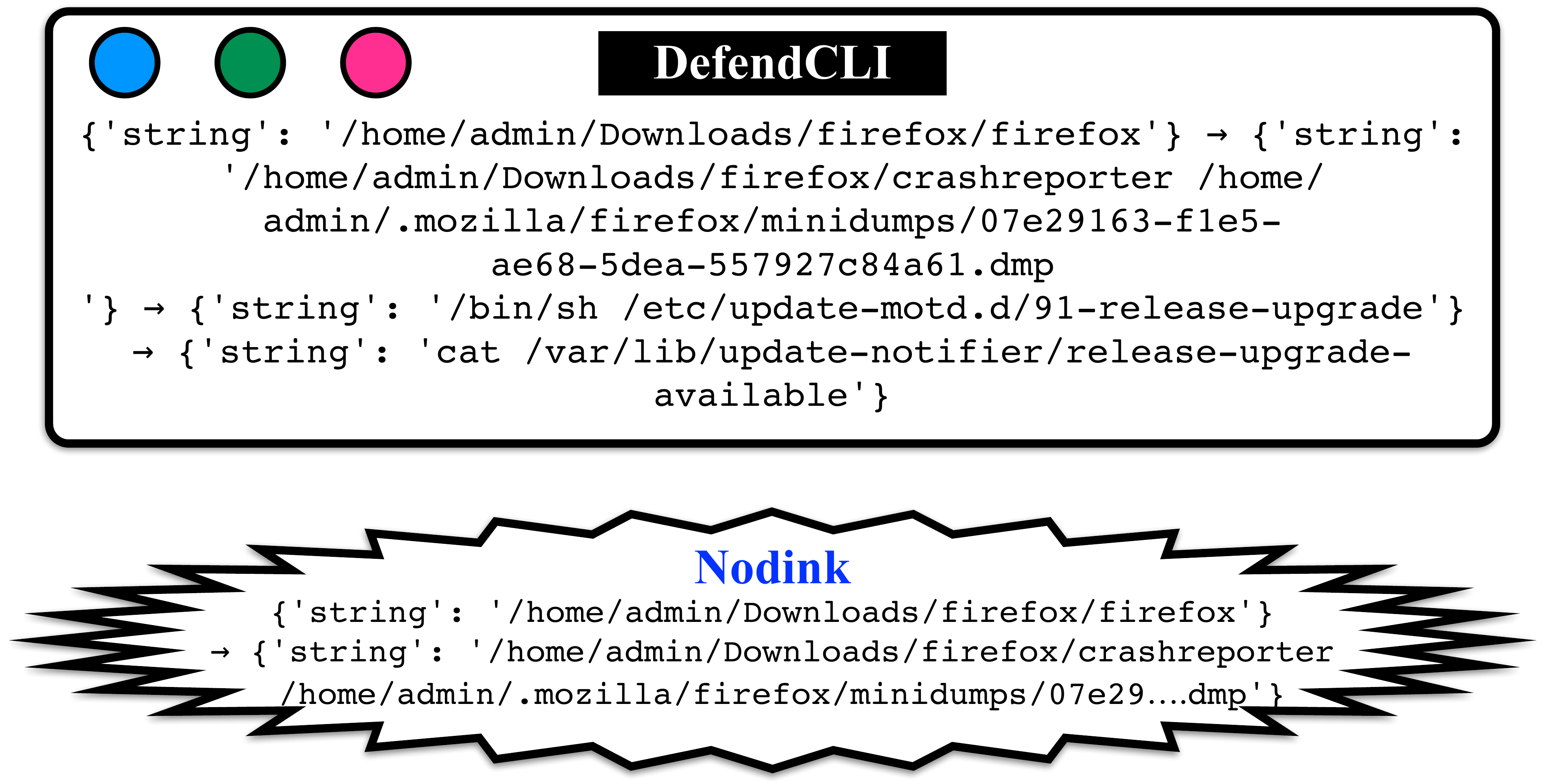}
        \caption{Firefox Extension Attack}
        \label{fig:firefox_extension}
    \end{subfigure}
    \caption{Visualization of the differences in detected command-line execution chains between \textsc{DefendCLI} and \textsc{Nodlink} \cite{li2023nodlink}, based on the E3-series datasets and real-time evaluation.}
    \label{fig:cmd_nodlink_defendcli}
\end{figure*}

%\begin{figure}[t]
%       % \vspace{10pt} % 调整行间距
%        \centering
%        \includegraphics[width=\linewidth]{v12.pdf}
%        \caption{Visualization of (1) Process Injection and Command-Line Evasion; (2) LOLBins with GPT recommender.}
%        \label{fig:v12}
%\end{figure}

\subsubsection{\textbf{Enhancing the Practicability of our Solution for Actionable Attack Provenance Analysis}}
We replace the traditional alarm scoring framework with a GPT-powered recommender system to prioritize critical alarms, effectively reducing alarm fatigue and improving response efficiency.
Figure \ref{fig:v12} shows the visualization of some InfoPaths that are detected in our real-time evaluation with the GPT recommender.
The GPT recommender first analyzes and explains the intent behind each command-line that appears potentially malicious, and then prioritizes the alarm for further examination.

Figure \ref{fig:gpt_results} presents the overall results of the GPT recommender across E3 datasets and real-time attack scenarios.
The GPT recommender generates the top-5 malicious snapshot alarms for initial examination, while the remaining alarms are restored for further review. If the top-5 snapshot alarms exhibit malicious traces, security analysts then proceed to examine the other alarms to gather additional evidence.

\subsection{The Improvement of Command-Line Detection Capability}
Table \ref{tab:cmd_scale} presents a comparison of the scale of attack-related command-line activities detected by \textsc{DefendCLI} and \textsc{Nodlink} \cite{li2023nodlink}.
It can be observed that our solution demonstrates superior overall effectiveness in identifying a greater number of malicious command-line activities.

\textbf{$\blacksquare$ How well does \textsc{DefendCLI} in the detection of obfuscation compared to \textsc{Nodlink}?}
Figures \ref{fig:transformation} and \ref{fig:obfuscation} illustrate the detection results of both detectors in handling advanced command-line transformations and Base64-encoded shell command obfuscation.

In Figure \ref{fig:transformation}, the string \(732F5B...202F67\) represents a transformation of an ASCII command into its hexadecimal form as a defense evasion technique.
When converted back to ASCII, it reveals the command \( s/[[:blank:]]\textbackslash{}+/ /g \), which is used to clean up injected spaces in obfuscated shellcode designed to evade detection.

In Figure \ref{fig:obfuscation}, the string \(2E2F67...6C2026\) represents a base64 obfuscated command-line for defense evasion.
When decoded the content, it reveals the command \( ./gtcache~\&\textgreater{} /dev/null~\& \), which runs a backdoor process in the background, suppressing output and ensuring persistent session activity without logging.

Based on our evaluation results, \textsc{DefendCLI} successfully detects advanced command-line attacks, whereas \textsc{Nodlink} \cite{li2023nodlink} fail to identify them.
This demonstrates the superior capability of \textsc{DefendCLI} in handling complex command-line obfuscation techniques, ensuring more comprehensive attack detection.

\textbf{$\blacksquare$ How well does \textsc{DefendCLI} in the attack correlation with the low-frequency event compared to \textsc{Nodlink}?}
Figures \ref{fig:tcexec}, \ref{fig:log_deletion}, \ref{fig:time_forgery}, and \ref{fig:firefox_extension} illustrate the detection results of both detectors in handling complex attack-chain correlations involving low-frequency events.

It can be observed that \textsc{DefendCLI} exhibits superior overall effectiveness in identifying and correlating a broader range of relevant command-line activities, improving the analysis of the provenance of the attack. 

In contrast, \textsc{Nodlink} \cite{li2023nodlink} fails to detect the command \(chmod~+x~tcexec\) for escalating the backdoor privileges of \(tcexec\) as shown in Figure \ref{fig:tcexec}.  
It also does not identify \(clear\_console\) as the root cause of the attack as shown in Figure \ref{fig:log_deletion} and is unable to reconstruct the full sequence of attack activities due to the absence of the timestamp forgery command \(date -d null +\%s\) as shown in Figure \ref{fig:time_forgery}.  
While it detects the crashed execution of the Firefox browser, it fails to correlate subsequent actions involving the \(update-mode\) script, which signifies a deeper attack footprint related to information modification as shown in Figure \ref{fig:firefox_extension}.

\subsection{Runtime Performance}
Table~\ref{tab:Runtime} presents the execution performance of \textsc{DefendCLI} across diverse datasets, highlighting its efficiency through parallelization. For the E3 data sets, \textsc{DefendCLI} runs on a 64,GB RAM, 16-core virtual machine, processing approximately 5,000 records in 6 seconds using multiprocessing. For industrial-scale, real-time detection tasks (A1, A2, A3), the system is deployed on Microsoft Azure DataLake~\cite{microsfot-datalake}, leveraging Kafka-based streaming and ETL pipelines. Under this setup, \textsc{DefendCLI} achieves end-to-end response times of roughly 2 seconds, including network latency, for high-volume threat detection workloads.

\begin{table}[t]
\caption{The Scale of Attack-Related Command-Line Detected by \textsc{DefendCLI} and \textsc{Nodlink}}
\centering
\begin{tabular}{@{}l|c|cc@{}}
\toprule
\textbf{Dataset} & \textbf{\textsc{DefendCLI}} & \textbf{\textsc{Nodlink}} &\textbf{Potential Weakness}\\
\midrule
E3-TRACE       & 18\% & 11\% & Obfuscation \\
\midrule
E3-THEIA       & 19\%  & 17\%  & Context Awareness \\
\midrule
E3-CADETS      & 18\%  & 13\% & Low-Frequency Event\\
\midrule
A1             & 15\%  & 8\%  & Context Awareness\\
\midrule
A2             & 25\%  & 12\% &  Low-Frequency Event \\
\midrule
A3             & 9\%   & 3\%  &  Low-Frequency Event \\
\bottomrule
\end{tabular}
\label{tab:cmd_scale}
\end{table}

\begin{table}[t]
\caption{Runtime Performance}
\centering
\begin{tabular}{@{}l|c|c|c@{}}
\toprule
\textbf{Dataset}     & \textbf{Runtime} & \textbf{Batch Size}& \textbf{Platform}
\\
\midrule
E3-TRACE & $\approx$ 5 secs & 5,000 records & 64GB RAM - 16 Cores\\
\midrule
E3-THEIA & $\approx$ 5 secs & 5,000 records & 64GB RAM - 16 Cores\\
\midrule
E3-CADETS & $\approx$ 7 secs  & 5,000 records & 64GB RAM - 16 Cores\\
\midrule
A1 A2 A3 & $\approx$ 2 secs & / & Microsoft Azure DataLake\\
\bottomrule
\end{tabular}
\label{tab:Runtime}
\end{table}

\section{Conclusion}
In conclusion, \textsc{DefendCLI} addresses critical gaps by introducing a multilevel command-line-driven differentiation framework that enhances attack provenance analysis.  
By leveraging graph algorithms, explainable AI, and multi-stage correlation, \textsc{DefendCLI} refines provenance analysis to detect attacks existing methods often miss. Unlike traditional models relying on pre-trained normal behavior profiles, it provides contextually aware anomaly detection, improving precision while reducing irrelevant alerts.  

Through an extensive evaluation of real-world APT threats, \textsc{DefendCLI} demonstrates effectiveness in identifying sophisticated attack chains, detecting overlooked threats, and improving actionable intelligence for security teams. By enhancing reliability, reducing detection noise, and increasing context awareness, it delivers a practical, scalable, high-precision EDR solution to strengthen cybersecurity defenses.

%\section*{Ethic Consideration}
%We have carefully considered the ethical implications of this research and believe that the work does not pose any ethical or legal risks. The study does not involve human subjects, personal or sensitive data, or the analysis or disclosure of vulnerabilities. No potentially harmful outcomes are foreseen as a result of conducting or publishing this research. Consequently, an ethics review (e.g., IRB approval) was not required. We affirm that the research adheres to accepted ethical standards for computer and information security research, as outlined in the Menlo Report.

{\linespread{1.2} \footnotesize\bibliographystyle{ieeetr}
\bibliography{reference}}

\end{document}